\documentclass[AMA,STIX1COL]{WileyNJD-v2}

\usepackage{mathtools,mathrsfs,amsmath}
\usepackage{graphicx,caption,subcaption,color}
\usepackage{soul}
\usepackage{lineno}

\DeclarePairedDelimiter\ceil{\lceil}{\rceil}

\articletype{Article Type}%

\received{26 April 2016}
\revised{6 June 2016}
\accepted{6 June 2016}

\newcommand{\smhec}{SMHEC}

\begin{document}

\title{Simultaneous Moving Horizon Estimation and Control for Nonlinear Systems Subject to Bounded Disturbances}

\author[1]{Nestor Deniz*}

\author[1]{Guido Sanchez}

\author[1]{Marina Murillo}

\author[1]{Leonardo Giovanini}

\authormark{Nestor Deniz \textsc{et al}}

\address[1]{\orgdiv{Instituto de Investigacion en Senales, Sistemas e Inteligencia Computacional}, \orgname{sinc(i),UNL-CONICET}, \orgaddress{\state{Santa Fe}, \country{Argentina}}}

\corres{*Nestor Deniz. \email{ndeniz@sinc.unl.edu.ar}}

\abstract[Summary]{In this work, we address the output--feedback control problem for nonlinear systems under bounded disturbances using a moving horizon approach. The controller is posed as an optimisation-based problem that simultaneously estimates the state trajectory and computes future control inputs. It minimises a criterion that involves finite backward and forward horizons with respect to the unknown initial state, measurement noises and control input variables.
\textcolor{blue}{The main novelty of this work relies on linking the lengths of the forward and backward windows with the closed-loop stability, assuming detectability and decoding sufficient conditions to assure system stabilizability. It leads to a formulation that does not require to be a Control Lyapunov Function for the terminal cost of the controller.
Simulation examples are carried out to compare the performance of solving simultaneously and independently the estimation and control problems. Furthermore, the examples show how the controller influences the length of the estimation window through its gain.
}}


\maketitle

\section{Introduction}\label{Introduction}

One of the most popular control techniques {both in} academia and industry is Model Predictive Control (\emph{MPC}) due to its ability to explicitly accommodate hard state and input constraints \cite{rawlings2017model}. Thereon, much effort has been made to develop a stability theory for \emph{MPC} \cite{rawlings2017model, grune2011nonlinear}. An overview of recent developments can be found in Rakovi{\'c} and Levine \cite{rakovic2018handbook}. \emph{MPC} involves the solution of an open--loop optimal control problem at each sampling time with the current state as the initial condition. Each of these optimisations provide sequences of future control actions and states. The first element of the control action sequence is applied to the system and the optimisation problem is solved again at the next sampling time after updating the initial condition with the system state. \emph{MPC} keeps the computational burden constant by optimising the system behaviour within a finite length window. The system behaviour beyond the window is summarised in a term known as \emph{cost--to--go}. 

\emph{MPC} is often formulated assuming that the system state can be measured. However, in many practical cases, the only information available {are} noisy measurements of system output, so the use of independent algorithms for state estimation (including observers, filters and estimators) becomes necessary \cite{rawlings2006particle}. From all of these methods, {M}oving {H}orizon {E}stimation (\emph{MHE}) is especially engaging for {its} use with \emph{MPC} because it can be formulated as a similar online optimisation problem. Solving the \emph{MHE} problem produces an estimated state that is compatible with a set of past measurements that recedes as current time advances \cite{schweppe1973uncertain, rao2001constrained, rao2003constrained}. This estimate is optimal in the sense that it maximises a criterion that captures the likelihood of the measurements. {Despite the fact} that relevant results on \emph{MPC} were developed, research works on \emph{MHE} {have just} begun. The works of Rao et al. \cite{rao2001constrained, rao2003constrained} provide overviews of linear and nonlinear \emph{MHE}. Recent results regarding \emph{MHE} for nonlinear systems are given for robust stability and estimate convergence properties by Alessandri et al. \cite{alessandri2008moving, alessandri2012min}, Garcia-Tirado et al. \cite{garcia2016new} and Sanchez et al. \cite{sanchez2017adaptive}, among others. In recent years, several results have been obtained for different \emph{MHE} formulations, advancing from idealistic assumptions, like observability and vanishing disturbances, to realistic situations like detectability and bounded disturbances \cite{muller2017nonlinear, allan2019lyapunov, deniz2019robust}. {\color{blue} Zou et al. \cite{zou2020moving} are concerned with the MHE problem for networked linear systems with unknown inputs under dynamic quantization effects. The authors present a two-step MHE design strategy, which they name the decoupling and convergence steps. They provide sufficient conditions under which the estimation error is bounded and illustrate the effectivenes of the proposed method with two simulation examples.}

When disturbances, model uncertainties and system constraints can be neglected, state and control sequences can be independently computed \cite{duncan1971solutions, bensoussan2004stochastic, georgiou2013separation}. However, in practical applications, these conditions are very difficult to fulfil, i.e., process disturbances and measurement noise are usually present, as well as model uncertainty. In this context, it becomes necessary to use an approach that includes this information into the controller design. State-feedback \emph{MPC} is a mature field with results that consider model uncertainty, input disturbances, and noises \cite{raimondo2009min}. However, these works did not consider robustness with respect to errors in state estimation. Fewer results are available for output-feedback \emph{MPC}. An overview of nonlinear output-feedback \emph{MPC} is given by Findeisen et al. \cite{findeisen2003state} and the references therein. Many of these approaches involve designing a separate estimator and controller \cite{patwardhan2012nonlinear, zhang2013lyapunov, ellis2017state}. Results on robust output-feedback \emph{MPC} for constrained, linear, discrete-time systems with bounded disturbances and measurement noise can be found in Mayne et al. \cite{mayne2006robust, mayne2009robust} and Voelker et al. \cite{voelker2010unconstrained, voelker2013moving}. These approaches first solve the estimation problem and prove the convergence of the estimated state to a bounded set, and then take the uncertainty of the estimation into account when solving the \emph{MPC} problem. 

\textcolor{blue}{All these works rely on the assumption that estimation and control problems can be solved separately. However, in a context where process disturbances and noisy measurements are present, solving both problems separately may lead to suboptimal solutions \cite{mitter1996filtering}. Moreover, stability analysis would be a complicated task when disturbances are not Gaussian. The problem becomes more challenging when the system is nonlinear, constraints on states and/or inputs are present, and the model is not accurate enough \cite{georgiou2013separation, kwakernaak1974linear}. Since most of these conditions are usual in real-world applications, analysing and solving the estimation and control problems in a simultaneous approach would seem the most suitable.
}

The approach of solving simultaneously \emph{MHE--MPC} was originally introduced by Copp and Hespanha \cite{copp2014nonlinear} and later developed \citep{copp2017simultaneous}. In the first paper, Copp and Hespanha \cite{copp2014nonlinear} proposed an output feedback controller that combines state estimation and control into a single \textit{min-max} optimisation problem that, under observability and controllability assumptions, guarantees the boundedness of state and tracking errors. Finally, in their last work, Copp and Hespanha \cite{copp2017simultaneous} established the conditions for guaranteeing the boundedness of error for trajectory tracking problems. They also introduced a primal--dual interior point method that can be used to efficiently solve the \textit{min-max} optimisation problem. The criterion used in these works involves finite forward and backward horizons that are minimised with respect to feedback control policies and maximised with respect to the unknown parameters in order to guaranty robustness in the worst-case scenario. \textcolor{blue}{The former results in simultaneous estimation and control {have} seen to overcome the difficulties that appear in real-world applications. However, physical interpretations are missed or hidden behind the \textit{min-max} formulation.}

\textcolor{blue}{In this article, we introduce an output--feedback controller for nonlinear systems subject to bounded disturbances using a simultaneous moving horizon estimation and control (\smhec{}) approach. This concept implies that we are solving at each iteration an estimation and control problem at the same time. The most common approach found in the literature solves the estimation problem first and then, the control one. In our work, the resulting optimisation problem minimises a criterion that involves finite backward and forward horizons with respect to the unknown initial state, measurement noise and control input variables while it is maximised with respect to the unknown future disturbance variables. The optimisation problem is stated as {a} minimisation of {an} objective function that {takes into account} the notions of detectability and controllability. The former is a system's property {that} the estimation window takes advantage {of} for estimating the optimal state. The latter relates to the ability of the controller to steer the system to the desired operation zone, given a set of controls and constraints over a finite-time forward window whilst exogenous and unknown inputs disturb the system.} 
We show that the proposed controller results in closed--loop trajectories along which the states remain bounded. These results rely on two assumptions: the first assumption requires that the optimisation criterion includes the adaptive arrival cost proposed by Sanchez et al. \cite{sanchez2017adaptive}. This assumption allows {us} to ensure the boundedness of the state estimate and to obtain a bound for the estimation error set if the parameters of the estimation problem are properly chosen \cite{deniz2019robust}.
The second assumption requires that the backward (estimation) and forward (control) horizons are sufficiently large so that enough information is obtained to find state estimates and control inputs compatible with the system dynamics, noises and constraints. This assumption is satisfied if the system is detectable, stabilizable and the parameters in the cost function (weights and horizons) are chosen appropriately.

Th{is} paper is organised as follows: the estimation and control problem --including notation, definitions and properties that will be used throughout the paper-- are introduced in Section \ref{section:preliminaries}. The analysis of closed-loop stability of the proposed algorithm is presented in Section \ref{section:robust}. In Section \ref{section:examples}, two examples to illustrate the concepts presented in this work are given. The first example uses a simple nonlinear model to analyse the consequences of simultaneously solving the estimation and control problems. The second example compares the performance obtained by the simultaneous and independent approaches applied to the regulation problem of the Van der Pol oscillator for two operational conditions. Finally, conclusions and future work are discussed in Section \ref{section:conclusion}.

\section{Preliminaries and setup} \label{section:preliminaries}
\subsection{Notation}

Let $\mathbb{Z}$ denote the integer numbers, $\mathbb{Z}_{\left[a,b \right]}$ denotes the set of integers in the interval $\left[ a,b \right]$, with $b>a$ and $\mathbb{Z}_{\geq a}$ denotes the set of integers greater or equal to $a$. Boldface symbols denote sequences of finite ($\boldsymbol{w} \coloneqq \{ w_1, \ldots, {w_p} \}$) or infinite ($\boldsymbol{w} \coloneqq \{ w_1, \ldots, {w_p}, \ldots \}$) length. We denote $\hat{x}_{j\vert k}$ as the state at time $j$ estimated at time $k$. 
By $\vert x\vert$ we denote the euclidean norm of a vector $x \in \mathbb{R}^{n_x}$. Let $\Vert \boldsymbol{x}\Vert \coloneqq \sup_{k\in \mathbb{Z}_{\geq 0}} \vert x_k\vert $ denote the supreme norm of the sequence $\boldsymbol{x}$ and $\Vert \boldsymbol{x}\Vert_{\left[a, b \right]} \coloneqq \sup_{k\in \mathbb{Z}_{\left[a, b \right]}} \vert x_k\vert$. A function $\gamma : \mathbb{R}_{\geq 0} \rightarrow \mathbb{R}_{\geq 0}$ is of class $\mathcal{K}$ if $\gamma$ is continuous, strictly increasing and $\gamma \left(0\right) = 0$ . If $\gamma$ is also unbounded, it is of class $\mathcal{K}_\infty$. A function $\zeta : \mathbb{R}_{\geq 0} \rightarrow \mathbb{R}_{\geq 0}$ is of class $\mathcal{L}$ if $\zeta$ is continuous, non increasing and $\lim_{t \rightarrow \infty} \zeta\left(t \right) = 0$. A function $\beta : \mathbb{R}_{\geq 0}\times\mathbb{Z}_{\geq 0} \rightarrow \mathbb{R}_{\geq 0}$ is of class $\mathcal{KL}$ if $\beta \left(\cdot,k \right)$ is of class $\mathcal{K}$ for each fixed $k \in \mathbb{Z}_{\geq 0}$, and $\beta \left(r,\cdot \right)$ of class $\mathcal{L}$ for each fixed $r \in \mathbb{R}_{\geq 0}$.
In the following sections, we will use the notation $\Psi_{\textnormal{\fontsize{6pt}{8pt}\selectfont\textit{p,t,l}}}$ to reference the cost incurred solving the problem \textit{p} at time \textit{t} with a horizon length \textit{l}, while $\Psi_{\textnormal{\fontsize{6pt}{8pt}\selectfont\textit{p,t,l}}}\left(x \right)$ will be used to indicate the cost at the solution $x$, with $x$ belonging to a consistent domain with the cost function $\Psi_{\textnormal{\fontsize{6pt}{8pt}\selectfont\textit{p,t,l}}}$. When necessary, we will use the notation $x^{(1)}_{i,k}$ and $x^{(2)}_{i,k}$ to differentiate $i$--th component of the state vector of two different discrete-time trajectories of the system, with $i\in\mathbb{Z}_{\left[1,n\right]}$. Moreover, $x^{(1)}_k(x_0^{(1)},\boldsymbol{w}^{(1)} )$ will denote a trajectory with initial condition $x_0^{(1)}$ and perturbed by the sequence $\boldsymbol{w}^{(1)}$. A similar notation is used for the case of continuous time systems, where $t$ is used instead $k$ to denote continuous time.

\subsection{Problem statement}

Let us consider a discrete-time nonlinear system whose behaviour is given

\begin{equation}        \label{nonlinear system}
  \begin{array}{rl}
    x_{k+1} =&    f\left(x_k,u_k\right)+w_k,  \quad \forall \, k \in \mathbb{Z}_{\geq 0}, \\
    y_k =& h\left(x_k \right) + v_k, 
    \end{array}
\end{equation}

\noindent where $x \in \mathcal{X} \subset \mathbb{R}^{n_x}$ is the system{'s} state, $u \in \mathcal{U} \subset \mathbb{R}^{n_u}$ is the system's input and $w \in \mathcal{W} \subset \mathbb{R}^{n_w}$ is the unmeasured process disturbance posed as an additive input. The output of the system is $y \in \mathcal{Y} \subset \mathbb{R}^{n_y}$ and $v \in \mathcal{V} \subset \mathbb{R}^{n_v}$ is the measurement noise. The function $f(\cdot,\cdot)$ is assumed to be at least locally Lipschitz in its arguments, and the function $h(\cdot)$ is known to be a continuous function.
The sets $\mathcal{X}$, $\mathcal{U}$, $\mathcal{W}$, $\mathcal{V}$ and $\mathcal{Y}$ are assumed to be convex, containing the origin in its interior.
The estimation and control problem attempts to simultaneously find the optimal state $\hat{x}_{k\vert k}$ and the optimal sequence of control inputs $\boldsymbol{\hat{u}}$ which steer the system to the desired operation zone. It is in an infinite-horizon optimisation problem given by

\begin{equation}        \label{mhe mpc infinite horizon}
  \begin{array}{l}
   \underset{\substack{\hat{x}_{0|k}, \boldsymbol{\hat{w}}, \boldsymbol{\hat{u}} } } {\operatorname{min}}  \Psi_{\textnormal{\fontsize{6pt}{8pt}\selectfont\textit{EC,k,$\infty$}}} \coloneqq \displaystyle \sum_{j=0}^{k} \ell_e \left(\Hat{w}_{j\vert k}, \hat{v}_{j\vert k} \right) + \sum_{j=k}^{\infty}\left(\ell_c \left(\hat{x}_{j\vert k}, \hat{u}_{j\vert k} \right) - \ell_{w_c}\left( \hat{w}_{j\vert k} \right)\right) \\
   \hspace{1cm} \text{s.t.} \left\{
   \begin{array}{l}
     \begin{array}{rl}
       \hat{x}_{j+1|k} =& f\left(\hat{x}_{j\vert k},\hat{u}_{j\vert k}\right)+\hat{w}_{j\vert k},        \\
        y_{j} =& h\left(\hat{x}_{j\vert k}\right) + \hat{v}_{j\vert k},\\  
     \end{array} \\
     \hat{x}_{j|k} \in \mathcal{X}
     ,\,\hat{u}_{j\vert k}\in \mathcal{U}, \hat{w}_{j|k} \in  \mathcal{W}, \, \hat{v}_{j|k} \in \mathcal{V}.
    \end{array} \right.
  \end{array}
\end{equation}

{\color{blue} where functions $\ell_e(\hat{w}_{j\vert k},\hat{v}_{j\vert k})$, $\ell_{c}(\hat{x}_{j\vert k},\hat{u}_{j\vert k})$ and $\ell_{w_c}(\hat{w}_{j\vert k})$ are norms. Therefore, it is clear that they penalise large values of their arguments within their windows ($\hat{w}_{j\vert k}$, $\hat{v}_{j\vert k}$ $\forall j \leq k$ {and} $\hat{x}_{j|k}$, $\hat{u}_{j|k}$ $\forall j \geq k$). However, the effect of $\ell_{w_c}(\hat{w}_{j\vert k})$ on the objective function $\Psi_{EC,k,\infty}$ is different: since $\ell_{w_c}(\hat{w}_{j\vert k})$ is substracting, it will maximise the value of $\hat{w}_{j|k}$ $\forall j \geq k$ given that $\hat{w}_{j|k} \in \mathcal{W}$. In this way, $\Psi_{EC,k,\infty}$ encodes a measure of robustness since, as it will be shown later, $\ell_{c}(\hat{x}_{j\vert k},\hat{u}_{j\vert k})$ and $\ell_{w_c}(\hat{w}_{j\vert k})$ are related to the ability of the system to mitigate deviations due to disturbances (actual and future ones) within the forward window.}

{\color{blue}Problem \eqref{mhe mpc infinite horizon} is valuable from a theoretical point of view, since it guarantees the boundness of estimates $\hat{x}_{j|k}$ and control inputs $\hat{u}_{j|k}$ provided that the cost function is bounded, $\Psi_{EC,k,\infty} < \gamma$ $\forall k \in \mathbb{Z}_{\geq 0}$, $\gamma \in \mathbb{R}_{\geq 0}$. In fact, if functions $\ell_e(\cdot,\cdot)$, $\ell_c(\cdot,\cdot)$ and $\ell_{w_c}(\cdot)$ are defined using a norm--$\ell_p$, then Problem \eqref{mhe mpc infinite horizon} would guarantee that $x_k$ and $u_k$ are $\ell_p$, provided that noises $w_k$ and $v_k$ are also $\ell_p$. This would mean that the closed-loop system has a finite $\ell_p$--induced gain and therefore stable. In this way, problem \eqref{mhe mpc infinite horizon} becomes a benchmark for any simultaneous estimation and control problem formulated for practical situations.
Since this problem uses an infinite--horizon, it is intractable from a computational point of view. Thus, it can be reformulated into a receding horizon problem with finite windows and extra terms that summarise the information outside the estimation and control windows. Following these concepts, problem \eqref{mhe mpc infinite horizon} can be rewritten as follows}

\begin{equation}        \label{eq_mhe mpc opt problem}
 \begin{array}{l}
  \displaystyle \underset{\substack{\hat{x}_{k-N_e|k}, 
    \boldsymbol{\hat{w}}, \boldsymbol{\hat{u}} } }
    {\operatorname{min}} \, \Psi_{\textnormal{\fontsize{6pt}{8pt}\selectfont\textit{EC,k,$N_e+N_c$}}} \coloneqq
     \Gamma_{k-N_e}\left(\chi\right) +  \displaystyle \sum_{j=k-N_e}^{k} \ell_e \left(\hat{w}_{j\vert k}, \hat{v}_{j\vert k} \right) +  \\
     \displaystyle
     \hspace{2.0cm}  \sum_{j=k}^{k+N_c-1}\left(\ell_{c} \left(\Hat{x}_{j\vert k}, \hat{u}_{j\vert k} \right) - \ell_{w_c}\left( \hat{w}_{j\vert k} \right)\right) + \Upsilon_{k+N_c}\left( \Xi \right) \vspace{0.1cm}\\
     \hspace{1cm} \text{s.t.} \left\{
     \begin{array}{l}
       \begin{array}{rl} 
         \chi =& \hat{x}_{k-N_e\vert k}-\bar{x}_{k-N_e},    \\
         \Xi =& \hat{x}_{k+N_c\vert k},                     \\
         \hat{x}_{j+1|k} =& f\left(\hat{x}_{j\vert k},\hat{u}_{j\vert k}\right)+\hat{w}_{j\vert k}, \\
          y_{j} =& h\left(\hat{x}_{j\vert k}\right) + \hat{v}_{j\vert k},\\  
       \end{array} \\
       \hat{x}_{j|k} \in \mathcal{X}, \,
       \hat{u}_{j\vert k}\in \mathcal{U}, \hat{w}_{j|k} \in \mathcal{W}, \ \hat{v}_{j|k} \in \mathcal{V}.
    \end{array} \right.
  \end{array}
\end{equation}

{\color{blue}In this new formulation, the infinite horizon cost function $\Psi_{EC,k,\infty}$ has been replaced by the finite horizon cost function $\Psi_{EC,k,N_e+N_c}$. It is composed by two finite windows; one backward and one forward associated with the estimation ($\Psi_{E,k,N_e}$) and control ($\Psi_{C,k,N_c}$) problems, respectively. Each of these windows include a finite number of terms to measure the performance of the corresponding problem and an extra term to summarise the information outside these windows. The backward window

\begin{align}
    \Psi_{E,k,N_e} = \Gamma_{k-N_e}\left(\chi\right) +  \sum_{j=k-N_e}^{k} \ell_e \left(\hat{w}_{j\vert k}, \hat{v}_{j\vert k} \right)
\end{align}
includes $N_e$ terms backward in time corresponding to the estimation stage-cost $\ell_e(\hat{w}_{j\vert k}, \hat{v}_{j\vert k})$ and the arrival cost $\Gamma_{k-N_e}(\chi)$. The arrival cost summarises the information left behind the estimation window by penalising the uncertainty in the initial state $\hat{x}_{k-N_e|k}$ \cite{rao2001constrained,rao2003constrained}. The forward window

\begin{align}
    \Psi_{C,k,N_c} = \sum_{j=k}^{k+N_c-1}\left(\ell_{c} \left(\Hat{x}_{j\vert k}, \hat{u}_{j\vert k} \right) - \ell_{w_c}\left( \hat{w}_{j\vert k} \right)\right) + \Upsilon_{k+N_c}\left( \Xi \right)
\end{align}
includes $N_c$ terms forward in time corresponding to the \textit{controller stage-cost}, $\ell_{c}(\hat{x}_{j\vert k}, \hat{u}_{j\vert k})$, the future disturbances stage-cost $\ell_{w_c}(\hat{w}_{j\vert k})$, and the \textit{cost-to-go} $\Upsilon_{k+N_c}(\Xi)$. The \textit{cost-to-go} summarises the behaviour of the system beyond the control window  penalising the deviation of the final state $\hat{x}_{k+N_c \vert k}$ while remaining within the set of terminal constraints $\mathcal{X}_f$, as is commonly used in \emph{MPC} \citep{rawlings2017model}.
}

The goal of problem \eqref{eq_mhe mpc opt problem} is to estimate the initial state $\hat{x}_{k-N_e \vert k}$ and disturbances $\hat{w}_{j \vert k}$ $j \in\mathbb{Z}_{\left[k-N_e,k-1\right]}$ such that an estimate $\hat{x}_{k|k}$ is obtained to compute the control inputs $\hat{u}_{j \vert k}$ $j\in\mathbb{Z}_{\left[k,k+N_c-1\right]}$ that drive the system states to the desired region. Therefore, there is no point on penalizing the control cost $\ell_c(\cdot,\,\cdot)$ along the estimation window. The variables $\hat{v}_{j \vert k}$ are uniquely determined by the remaining optimisation variables and the output equation

\begin{align}
    \hat{v}_{j \vert k}  = y_j - h(\hat{x}_{j \vert k}), \qquad j \in\mathbb{Z}_{\left[k-N_e,k\right]}.
\end{align}

{\color{blue} Since future system outputs $y_j \; \forall j \in \mathbb{Z}_{>k}$ are not available at sample $k$, then the measurement noise $\hat{v}_{j|k} \; \forall j \in \mathbb{Z}_{>k}$ will not be included into the control function $\Psi_{C,k,N_c}$. On the other hand, disturbances $\hat{w}_{j|k}$ must be included in both windows ($\Psi_{E,k,N_e}$ and $\Psi_{C,k,N_c}$), because they are needed to solve problem \eqref{eq_mhe mpc opt problem}: \textbf{past disturbances} $\hat{w}_{j|k}$ with $j \in \mathbb{Z}_{\left[k-N_e, k\right]}$ are required to compute $\hat{x}_{k|k}$, which is used as initial condition of the control problem, while the \textbf{future disturbances} $\hat{w}_{j|k}$ with $j \in \mathbb{Z}_{\left[k, k+N_c\right]}$ are required to compute the control sequence $\mathbf{\hat{u}_{j|k}}$ that drive the system states to $\mathcal{X}_f$. Future disturbances are frequently unknown, thus they must be computed in a way that closed-loop stability and feasibility are guaranteed for bounded disturbances.

There are many ways of computing future disturbances \cite{rawlings2017model}. One way is computing $\hat{w}_{j|k}$ with $j \in \mathbb{Z}_{\left[k, k+N_c\right]}$ for the worst case scenario, in this case by maximising $\hat{w}_{j|k}$ for each sample time. Other way of computing future disturbances is through model augmentation \cite{pannocchia2015offset}. In this approach the disturbance is computed recursively through the system model, and it is removed from the optimisation problem. In this work, we will keep the disturbance term $\ell_{w_c}( \hat{w}_{j | k})$ within the control cost function $\Psi_{C,k,N_c}$ along the theoretical analysis to include the effects of disturbances on stability. For implementation, the future disturbances will be computed through model augmentation and estimation in order to reduce the computational burden of the optimisation problem.
}

\begin{remark}
The sequence of process disturbances $\hat{w}_{j\vert k}$ is minimised within the {estimation} window, i.e., $j\in \left[k-N_e-1,k-1\right]$, and it is maximised within the control window, $j\in \left[k,k+N_c-1\right]$.
\end{remark}

\subsection{Relationship with MHE and MPC}
The criterion $\Psi_{EC,k,N_e+N_c}$ can be rewritten as follows

\begin{equation}    \label{MHE+MPC multi_obj}
     \Psi_{EC,k,N_e+N_c} \coloneqq \varphi \Psi_{E,k,N_e}+(1-\varphi) \Psi_{C,k,N_c}, \quad \varphi \in \mathbb{R}_{[0,1]},
\end{equation}

which corresponds to a multi-objective formulation of $\Psi_{EC,k,N_e + N_c}$, where $\varphi$ controls the influence of $\Psi_{E,k,N_e}$ and $\Psi_{C,k,N_c}$. When $\varphi=0$, $\Psi_{EC,k,N_e+N_c} \coloneqq \Psi_{C,k,N_c}$ and problem \eqref{eq_mhe mpc opt problem} becomes a \emph{model predictive control} problem with terminal cost\cite{chen1998quasi}. On the other case, when $\varphi=1$, $\Psi_{EC,k,N_e+N_c} \coloneqq \Psi_{E,k,N_e}$ and problem \eqref{eq_mhe mpc opt problem} becomes a \emph{moving horizon estimation} problem\cite{garcia2016new, muller2017nonlinear, deniz2019robust, ji2016robust}. In these cases, the optimisation problem \eqref{eq_mhe mpc opt problem} has only one objective {function} and the separation principle needs to be applied since the estimator and the controller are implemented independently.

When $0 < \varphi < 1$, $\Psi_{E,k,N_e}$ and $\Psi_{C,k,N_c}$ are simultaneously considered by $\Psi_{EC,k,N_e+N_c}$ and the optimisation problem \eqref{eq_mhe mpc opt problem} becomes {a} multi-objective {one}. The relevance of $\Psi_{E,k,N_e}$, and therefore of $\Psi_{C,k,N_c}$, is defined by $\varphi$ emphasising or deemphasizing the influence of the estimation problem on the solution. In the case of $\varphi=0.5$, $\Psi_{E,k,N_e}$ and $\Psi_{C,k,N_c}$ have {a} similar influence on the solution of \eqref{eq_mhe mpc opt problem} and it becomes the problem proposed by Copp and Hespanha\cite{copp2017simultaneous}.

\begin{definition}
Given the points $z_E \in \mathcal{Z}_E \coloneqq \mathbb{R}^{n_w N_e}\times \mathbb{R}^{n_v(N_e+1)}\times \mathbb{R}^{n_x(N_e+1)}$ and $z_C \in \mathcal{Z}_C \coloneqq \mathbb{R}^{n_w N_c}\times \mathbb{R}^{n_u N_c}\times \mathbb{R}^{n_x(N_c+1)}$ such that $z \in \mathcal{Z} \coloneqq \mathcal{Z}_E \times \mathcal{Z}_C$. A point $z^{o} \in \mathcal{Z}$ is Pareto optimal if and only if there does not exist another point $z \in \mathcal{Z}$ such that $\Psi_{\textnormal{\fontsize{6pt}{8pt}\selectfont\textit{EC,$N_e+N_c$,k}}}(z) \leq \Psi_{\textnormal{\fontsize{6pt}{8pt}\selectfont\textit{EC,$N_e+N_c$,k}}}(z^o)$ and $\Psi_{\textnormal{\fontsize{6pt}{8pt}\selectfont\textit{E,$N_e$,k}}}(z_E)<\Psi_{\textnormal{\fontsize{6pt}{8pt}\selectfont\textit{E,$N_e$,k}}}(z_E^o)$, $\Psi_{\textnormal{\fontsize{6pt}{8pt}\selectfont\textit{C,$N_c$,k}}}(z_C)<\Psi_{\textnormal{\fontsize{6pt}{8pt}\selectfont\textit{C,$N_c$,k}}}(z_C^o)$ \citep{miettinen2012nonlinear}.
\label{def:def_1}
\end{definition}

According to {Definition \ref{def:def_1},} problem \eqref{eq_mhe mpc opt problem} looks for solutions {in a sense} that neither $\Psi_{E,k,N_e}$ nor $\Psi_{C,k,N_c}$ can be improved without {deteriorating} one of them. Any optimal solution of problem \eqref{eq_mhe mpc opt problem} 
with $0 < \varphi < 1$ is Pareto optimal \citep{miettinen2012nonlinear}, therefore it has an optimal trade-off between $\Psi_{E,k,N_e}$ and $\Psi_{C,k,N_c}$. {When} $\varphi=0$ or $\varphi=1$ the solutions of problem \eqref{eq_mhe mpc opt problem} are optimal in the sense of the selected objective {function} ($\Psi_{C,k,N_e}$ or $\Psi_{C,k,N_c}$, respectively). In {both} cases, the solutions obtained are not Pareto optimal and, therefore the overall system performance can be poorer than the one provided by the multi-objective problem.

From a practical point of view, $\varphi$ can be used to improve the numerical properties of the optimisation problem (\ref{eq_mhe mpc opt problem}). This fact allows to improve the convergence properties of numerical algorithms employed to solve it (see Example 4.2). For example, if $N_e \ll N_c$ and the stage costs $\ell_{e} \left(\cdot \right), \ell_{c} \left(\cdot \right)$ and $\ell_{w_c} \left(\cdot \right)$ have similar values, the optimisation problem will improve $\Psi_{C,k,N_c}$ at the expense of $\Psi_{E,k,N_e}$ (because $\Psi_{C,k,N_c} \gg \Psi_{E,k,N_e}$), deteriorating the estimation of $\hat{x}_{k|k}$ and{, thus,} producing potentially ill conditioned Jacobian and Hessian matrices of $\Psi_{EC,k,N_e+N_c}$. This numerical problems can lead to an increment of the computational {cost} of the optimisation problem. A similar situation can happen when $N_e \gg N_c$.

%% -------------------------------------------------------------------------
\section{Robust stability under bounded disturbances} \label{section:robust}

In this section, we introduce the results regarding feasibility and stability of the proposed algorithm. Firstly, the properties of \emph{MHE} and \emph{MPC} are analyzed and then the results for the simultaneous \emph{MHE--MPC} are given. Besides, feasibility conditions for the existence of a solution to problem \eqref{eq_mhe mpc opt problem} and minimum horizon lengths required to achieve the desired estimation and control performances are {also} analysed.

\subsection{Estimation problem}

{\color{blue}The proposed algorithm relies on the backward window to formulate an optimisation problem of fixed length $N_e$ to compute the optimal state estimate $\hat{x}_{k\vert k}$, which is used by the forward window (control problem) to compute $\hat{u}_{j|k}$. The estimation problem --and therefore problem \eqref{eq_mhe mpc opt problem}-- has a feasible solution if and only if the system is detectable. For non-linear systems, \textit{incremental input-output-to-state stability}  - \textit{i-IOSS}- is a popular characterization of detectability\cite{sontag1997output}

\begin{equation}    \label{eq: definicion i-IOSS}
  \begin{array}{rl}
    \vert x_k(x_0^{(1)},\boldsymbol{w}^{(1)} ) - x_k(x_0^{(2)},\boldsymbol{w}^{(2)})\vert &\leq
    \beta\left( \vert x_0^{(1)} - x_0^{(2)}\vert,k \right) + \gamma_1\left( \Vert\boldsymbol{w}^{(1)} -\boldsymbol{w}^{(2)}\Vert \right) 
    + \gamma_2\left( \Vert h\left(\boldsymbol{x}^{(1)} \right) -h\left(\boldsymbol{x}^{(2)} \right)\Vert \right), 
  \end{array}
\end{equation}
where $\beta(\cdot,\,\cdot)\in\mathcal{KL}$, $\gamma_1(\cdot),\,\gamma_2(\cdot)\in\mathcal{K}$. \textit{i-IOSS} not only guarantees the detectability of the system, but also means the system admits a robust stable state estimator \cite{allan2021nonlinear}. Note that inequality \eqref{eq: definicion i-IOSS} only includes the process disturbance as input to the system, however it can be extended to include control inputs. Moreover, as will be shown in Example \ref{example: 1}, the resulting control law have not only affects on the forward window but also in the backward window, influencing the estimation process.
}

Previous results on robust output-feedback \emph{MPC} with bounded disturbances the estimation problem is solved in the first place, showing the convergence of estimated states to a bounded set, and then take into account the uncertainty of estimation when solving the \emph{MPC} problem \citep{mayne2006robust, mayne2009robust}. The  key idea in these works was to consider the estimation error as an additional, unknown but bounded uncertainty that must be considered in order to guarantee the stability and feasibility of the resulting closed--loop system. {\color{blue}Let us define the robust estimable set 

\begin{equation}
  \mathcal{E}_{N_e} \left(\hat{x}_{k\vert k},\varepsilon_{e}(k) \right) \coloneqq \left\{x_k : \vert x_k-\hat{x}_{k\vert k} \vert \leq \varepsilon_{e}(k), \forall\, k  \right\},
\end{equation}
where $x_k$ is the real state, $\hat{x}_{k\vert k}$ is the best estimate of $x_k$ and $\varepsilon_e(k)$ is the estimation error at time bounded by

\begin{equation} \label{eq: bound estimation error}
    \varepsilon_e(k) \leq \bar{\Phi}\left( \vert x_0-\bar{x}_0\vert,k \right)+\pi_w\left(\Vert\boldsymbol{w}\Vert\right)+\pi_v\left(\Vert\boldsymbol{v}\Vert\right).
\end{equation}

This is a property of \textit{i-IOSS} systems\cite{sontag1997output}, where the parameters and structure of functions $\bar{\Phi}(\cdot)$, $\pi_w(\cdot)$ and $\pi_v(\cdot)$ depend on the stage costs $\ell_e(\hat{w}_{j|k},\hat{v}_{j|k})$ and arrival cost $\Gamma_{k-N_e}(\mathcal{X})$ employed in problem \eqref{eq_mhe mpc opt problem}, and they should be derived for each choice of  $\ell_e(\cdot)$ and $\Gamma_{k-N_e}(\cdot)$\cite{muller2017nonlinear,allan2019lyapunov,deniz2019robust}. In this work, we use the adaptive arrival cost proposed by Sanchez et al.\cite{sanchez2017adaptive}

\begin{align}
    \Gamma_{k - N_e}(\mathcal{X}) = \Vert \hat{x}_{k-Ne|k} - \hat{x}_{k-Ne+1|k-1} \Vert_{P^{-1}_{k-N_e|k}}
    \label{eq:arrival_cost_sanchez}
\end{align}
where $P^{-1}_{k-N_e|k}$ is updated using a recursive least square algorithm with variable forgetting factor\cite{sanchez2017adaptive}. This choice of terminal cost leads to the following functions\cite{deniz2019robust}

\begin{align}
  \bar{\Phi}\left(\vert x_0-\bar{x}_0 \vert, k\right) \coloneqq&\, \theta^i\vert x_0-\bar{x}_0 \vert^{\zeta} \frac{\mathbb{N}_e}{N_e}\left( \left(\frac{\overline{\lambda}_{P^{-1}} }{\underline{\lambda}_{P^{-1}}}\right)^{\rho}\left(c_{\beta}18^p + 
  \underline{\lambda}_{P^{-1}}^{\alpha_1}\left(P_{k-N_e}^{-1}  \right)\left(c_1\;3^{\alpha_1}+ c_2\;3^{\alpha_2}\right)\right) + c_{\beta}\;2^p\right), \label{eq: Phi} \vspace{2mm} \\
  \pi_w\left(\Vert\boldsymbol{w}\Vert \right) \coloneqq&\, 2\left(1+\mu\right)\left(\frac{c_{\beta}\;18^p}{\underline{\lambda}_{P^{-1}}} \;\bar{\gamma}_w^{\frac{p}{a}}\left(\Vert\boldsymbol{w}\Vert \right) +  c_2\; 3^{\alpha_2} \bar{\gamma}_w^{\alpha_2}\left(\Vert\boldsymbol{w}\Vert \right) + 
  \gamma_1\left(6 \Vert\boldsymbol{w}\Vert\right) + \gamma_1 \left(6 \underline{\gamma}_w^{-1} \left( 3 \bar{\gamma}_w\left(\Vert\boldsymbol{w}\Vert \right) \right) \right)\right),\label{eq: pi_w} \vspace{2mm} \\ 
  \pi_v\left(\Vert\boldsymbol{v}\Vert \right) \coloneqq&\, 2\left(1+\mu\right)\left(\frac{c_{\beta}\;18^p}{\underline{\lambda}_{P^{-1}}} \;\bar{\gamma}_v^{\frac{p}{a}}\left(\Vert\boldsymbol{v}\Vert \right) + c_1\; 3^{\alpha_1} \bar{\gamma}_v^{\alpha_1}\left(\Vert\boldsymbol{v}\Vert \right) + 
  \gamma_2\left( 6\Vert\boldsymbol{v}\Vert\right) + \gamma_2\left( 6\underline{\gamma}_v^{-1}\left( 3 \bar{\gamma}_v\left(\Vert\boldsymbol{v}\Vert \right) \right) \right)\right),\label{eq: pi_v}
\end{align}
where $\theta = \frac{2+\mu}{2(1+\mu)}<1$, $\mu \in \mathbb{R}_{\geq 0}$, $i=\lfloor \frac{k}{N_e}\rfloor$, $\rho=max\{p,\alpha_1,\alpha_2\}$, $\underline{\lambda}_{P^{-1}}$ and $\overline{\lambda}_{P^{-1}}$ are the minimal and maximal eigenvalues of $P^{-1}$.
The remaining parameters of equations \eqref{eq: Phi} to \eqref{eq: pi_v} are related to stage cost  $\ell_e(\hat{w}_{j|k},\hat{v}_{j|k})$ used in $\Psi_{E,k,N_e}$ and functions $\gamma_1(\cdot)$ and $\gamma_2(\cdot)$ of \textit{i-IOSS} definition. In fact, these are functions that bound the behaviour of $\ell_e(\hat{w}_{j|k},\hat{v}_{j|k})$ over time, thus $\gamma_1(\cdot)$ and $\gamma_2(\cdot)$ are related to $\ell_e(\hat{w}_{j|k},\hat{v}_{j|k})$\cite{deniz2019robust}. The key reason to use the arrival cost given by \eqref{eq:arrival_cost_sanchez} is the fact that the observer gains $\pi_w$ and $\pi_v$ only depend on the stage cost $\ell_e(\hat{w}_{j|k},\hat{v}_{j|k})$ and the properties of $w$ and $v$. 

Assuming an additive cost structure for the estimator

\begin{equation}
    \ell_{e}(\hat{w}_{j/k},\hat{v}_{j/k}) = \ell_{w_e}(\hat{w}_{j/k}) + \ell_{v_e}(\hat{v}_{j/k}),
\end{equation}
and quadratic functions for them, the functions involved in $\pi_w$ and $\pi_v$ get the following form

\begin{align}
 \gamma_1(\cdot)\leq & c_1 w^{\alpha_1} &
 \bar{\gamma}_w(\cdot)\leq & \bar{\lambda}_{Q_E} w^{2} \\
 \gamma_2(\cdot)\leq & c_2 v^{\alpha_2} &
 \bar{\gamma}_v(\cdot)\leq & \bar{\lambda}_{R_E} v^{2}   
\end{align}
where $c_1, c_2, \alpha_1, \alpha_2 \in \mathbb{R}_{\geq 0}$, $\bar{\lambda}_{Q_E}, \bar{\lambda}_{R_E}$ are the maximum and minimum eigenvalues of matrices $Q_E$ and $R_E$ respectively. and $N_e$ is the length of the backward window. $\mathbb{N}_e$ is the minimum length of the backward window required to guarantee the boundness of the estimation error, which is given by

\begin{equation}   \label{eq:minimo N}
  \mathbb{N}_e = \ceil*{ \left(2^{\rho} e_{max}^{\rho-1} \bar{c}_{\beta} \right)^{\frac{1}{\eta}}},
\end{equation}
where $\eta\in\mathbb{R}_{[0, 1/p]]}$ and $e_{\max}$ denotes the maximal error on the prior estimate of the initial condition. In the following, we will assume that $N_e\geq \mathbb{N}_e$.
}

At each sampling time, the measurements available along the backward window are used to obtain $\hat{x}_{k\vert k}$. Whenever $N_e \geq \mathbb{N}_e$, the estimation error will decrease until it reaches an invariant space whose volume depends on the process and measurement noises as well as the stage-cost and the system itself. The behaviour of the system is forecast\textcolor{blue}{ed} from the estimate $\hat{x}_{k\vert k}$, whereas $x_k$ remains within $\mathcal{E}_{N_e}$.

\subsection{Control problem}

\textcolor{blue}{The forward window corresponds to the receding horizon problem, which computes the optimal control input sequence $\boldsymbol{\hat{u}}$ using the optimal estimate $\hat{x}_{k\vert k}$, computed by the backward window, as the initial condition of the control problem. }
Its feasibility depends on the fact that its initial condition $x_{k}$ must belong to the \textit{robust $N_c$-steps controllable set} $\mathcal{R}_{N_c} \left(\Omega,\mathbb{T}\right)$ \citep{kerrigan2000invariant}

\begin{align}   \label{eq: conjunto robusto controlable}
    \mathcal{R}_{N_c}\left(\Omega,\mathbb{T}\right) \coloneqq& \left\{x_0\in\Omega|\exists\, u_{j}\in\mathcal{U}\,:\left\{x_{j}\in\Omega, x_{N_c}\in\mathbb{T}\right\} \quad \forall j \in \mathbb{Z}_{\left[0,N_c-1\right]} \right\}.
\end{align}
{\color{blue}where $\Omega$ is the input admissible set and $\mathbb{T}=\mathcal{X}_f$ is the terminal set of the controller.  Since $x_k \in \mathcal{E}_{N_e} \left(\hat{x}_{k\vert k},\varepsilon_{e} \right) \forall k$ the feasibility of the control problem is guaranteed if

\begin{equation}
  \mathcal{E}_{N_e} \left(\hat{x}_{k\vert k},\varepsilon_{e} \right) \subseteq \mathcal{R}_{N_c} \left(\Omega,\mathbb{T}\right) \quad \forall k\in \mathbb{R}_{\geq 0}.
\end{equation}
Note that this feasibility condition is not only necessary for the simultaneous \emph{MHE--MPC}, but also for independent \emph{MHE} and \emph{MPC} \citep{mayne2006robust,mayne2009robust}. Let us state this condition in the following assumption.}

\begin{assumption}      \label{assumption: feasibility condition}
The robust estimable set $\mathcal{E}_{N_e}$ belongs to the robust controllable set $\mathcal{R}_{N_c}\left(\Omega,\mathbb{T} \right)$ in $N_c$ steps $\forall k\in \mathbb{R}_{\geq 0}$, {$\mathcal{E}_{N_e} \subseteq \Omega,\, \mathcal{X}_f\subseteq\mathbb{T} $}, \textcolor{blue}{then, we are interested in the following robust controllable set:}{$\quad \mathcal{R}_{N_c}\left(\mathcal{E}_{N_e},\mathcal{X}_f\right) \quad \forall k\in \mathbb{R}_{\geq 0}$}.
\end{assumption}

\begin{remark}
\textcolor{blue}{This assumption formalises the compatibility notion between the estimator and the controller. Since the true system's state $x_k$ is not known perfectly, we require that it belongs to a neighbourhood of estimated one, $\hat{x}_{k\vert k}$, such that the control $u_k$ of the sequence $\boldsymbol{u}_{[k,k+N_c-1]}$ computed from $\hat{x}_{k\vert k}$ and applied to $f(x_k,u_k)$ will lead to the state $x_{x+1}$ not too far from the desired reference.
Moreover, the volume of the robust estimable set decrease faster with longer backward windows and the size of the robust controllable set can be enlarged by mean of larger forward window and with the appropriate design of the set $\mathcal{U}$.}
\end{remark}

\textcolor{blue}{The satisfaction of Assumption \ref{assumption: feasibility condition} can be easily verified since its depends on the adequate selection of problem (\ref{eq_mhe mpc opt problem}) parameters. This fact implies the selection of the estimation and control stage costs parameters ($Q_E, R_E, Q_C, R_C$ for quadratic cost functions), and the estimation and control horizons ($N_e, N_c$).}

Regarding stability along the forward window, a common approach to guarantee the stability of \emph{MPC} is by mean of the inclusion of a terminal constraint set, which is generally a level set of a control Lyapunov function \citep{mayne2000constrained}. This set is an artificial constraint set but {it} guarantees stability \citep{tuna2006shorter}. In this work we will analyse the stability of the controller following a similar approach as in Tuna, Messina and Teel\cite{tuna2006shorter}, where the analysis is carried out as a function of the length of the forward window, taking into account the effect of the process disturbances and the estimation errors.
A pseudo measure of the system controllability property will be introduced and the minimum forward window length which guarantees the stability of the simultaneous \emph{MHE--MPC} is given, without imposing {neither} extra terminal constraints nor {forcing} the cost-to-go to be a closed--loop Lyapunov function (\emph{CLF}). In this sense, let us state the following assumption.

\begin{assumption}  \label{assump relaxed contraction of Vf}
There exist a constant $\delta\in \mathbb{R}_{\geq 0}$ such that the cost-to-go and the stage cost satisfy the following relation:
\begin{equation}    \label{relaxed contraction of Vf}
    \Upsilon_{k+N_c}\left( f\left(x,u\right)\right)-\Upsilon_{k+N_c}\left( x \right)(1+\delta) \leq -\ell_{c}\left( x, u \right) + \ell_{w_c}\left( w \right).
  \end{equation}
\end{assumption}

\begin{remark}
\textcolor{blue}{A similar assumption was already used in Tuna, Messina and Teel\cite{tuna2006shorter}, where the constant $\delta$ is introduced in order to relax the requirement on function $\Upsilon_{k+N_c}\left( \cdot \right)$ to be a \emph{CLF} for the nominal case. }
\end{remark}

Despite {using} a different notation for the cost-to-go term $\Upsilon_{k+N_c}\left(\Xi\right)$, this function can take the same behaviour as the stage-cost, i.e., $\Upsilon_{k+N_c}\left(\Xi\right)=\ell_{c}\left(\Xi,0\right)$.
Here we extend it to {a} more general case where process disturbances are affecting the system, and it will lead, as {it} will be shown later, in longer control windows. 
Regarding the elements of the optimisation problem corresponding to the control problem, we will assume that the stage-cost is lower bounded.

\begin{assumption}      \label{assumption stage cost controller}
The stage cost $\ell_{c}\left( x,u \right)$ is lower bounded by a function $\sigma\left( x \right) \in \mathcal{K}_{\infty}$, such that $\sigma\left( x \right) \leq \ell_{c}\left( x,u \right) \quad \forall \, x \in \mathcal{X}, u \in \mathcal{U}$.
\end{assumption}

\begin{remark}
\textcolor{blue}{
Note that for a quadratic stage-cost, i.e., $\ell_{c}\left(x,u\right)=x^TQ_Cx+u^TR_Cu$, with $Q_C \succcurlyeq 0$ and $R_C \succ 0$, Assumption \ref{assumption stage cost controller} holds with $\sigma(x)=\underline{\lambda}_{Q_C}\vert x \vert^2$, where $\underline{\lambda}_{Q_C}$ denotes the minimal eigenvalue of matrix $Q_C$.}
\end{remark}

Furthermore, we will assume that there {is} an increasing sequence that relates the function $\sigma(x)$ with the cost of the control problem $\Psi_{C,k,i}$, where $i$ represents different lengths of the forward window. 
Finally, let us define the following quantity

\begin{align}
    \Delta^w_c\coloneqq
    \displaystyle \max \,
    \left\{\underset{\substack{\hat{u}_{k\vert k}} }{\operatorname{min:}\,}\frac{\ell_{w_c}\left(\hat{w}_{k\vert k}\right)}{\ell_c\left( \hat{x}_{k\vert k}, \hat{u}_{k\vert k} \right)}\right\},\,\forall\,\hat{x}_{k\vert k}\in\mathcal{X},\,\forall\,\hat{w}_{k\vert k}\in\mathcal{W}.
\end{align}
It encodes a pseudo--measure of the system controllability relating the capability of control actions to compensate process disturbances. The term pseudo--measure is used here because the relation $\Delta^w_c$ is given via the penalisation functions $\ell_{w_c}\left(\cdot\right)$ and $\ell_c\left(\cdot,\cdot\right)$. In the following, we will assume that the system is controllable from this point of view.

\begin{assumption}  \label{assumption:condition for stability}
The system's actuator can be designed such that the following relation can be always verified
\begin{align}
    \Delta^w_c <& 1
\end{align}

\textcolor{blue}{
\begin{remark}
It entails that the system's actuator is able to compensate for deviation of the system's output due to disturbances. Roughly speaking, the actuator is able to generate control actions able to compensate the effect of the disturbances. This fact is equivalent to require a system bandwidth wider than the disturbance ones.
\end{remark}}
\end{assumption}

Finally, let us state the following common assumption that establish lower and upper bounds on the cost function
\begin{assumption}      \label{bound of partial cost}
There exists an increasing sequence $\boldsymbol{L}\coloneqq\left[1,L_1,\ldots,L_i\right]$, $L_i \in \mathbb{R}$, $1\leq L_i \leq L$, $i \in\mathbb{Z}_{\geq 0}$ that verifies

\begin{equation}    \label{Ineq_A4}
  \sigma\left(x\right)\leq \Psi_{\textnormal{\fontsize{6pt}{8pt}\selectfont\textit{C,k,i}}} \leq  L_{i}\,\sigma\left(x\right).   
\end{equation}
\end{assumption}

\begin{remark}
\textcolor{blue}{
Note that the behaviour of the sequence $\boldsymbol{L}$ unveils the stability (or instability) associated with sequence $\Psi_{C,k,i} \quad i \in \mathbb{Z}_{\geq 0}$. If the increasing sequence $\boldsymbol{L}$ is bounded, then $\Psi_{C,k,N_c}$ will be bounded and the resulting closed-loop system stable.}
\end{remark}

\subsection{Simultaneous estimation and control}
With all the elements introduced in the previous section, we are ready to derive the main result: the stability of the resulting closed-loop system of the proposed output-feedback controller with estimation horizon $\mathbb{N}_e$ and control horizon $\mathbb{N}_c$ for nonlinear detectable and controllable systems under bounded disturbances. 

\begin{theorem}     \label{theorem: forward window}
Given the \textit{i}-\emph{IOSS} nonlinear system \eqref{nonlinear system} with a prior estimate $\bar{x}_0 \in \mathcal{X}_0$ of its unknown initial condition $x_0$, bounded disturbances $\boldsymbol{w} \in \mathcal{W}\left(w_{\max} \right)$, $\boldsymbol{v} \in \mathcal{V}\left(v_{\max} \right)$, Assumptions \ref{assumption: feasibility condition} to \ref{assumption:condition for stability} are fulfilled \textcolor{blue}{and the estimation and control horizons verify that}

\begin{align}        \label{eq:stabilizinc lenght contrl window}
    N_e \geq \mathbb{N}_e = \ceil*{ \left(2^{\zeta} e_{max}^{\zeta-1} \bar{c}_{\beta} \right)^{\frac{1}{\eta}}},\qquad N_c \geq \mathbb{N}_c = \ceil*{\displaystyle 1+\frac{\ln{\left(\frac{\delta\left(L-1\right)}{1-\Delta^w_c}\right)}}{\ln{\left(\frac{L}{L-1}\right)}}},
\end{align}
\textcolor{blue}{
then the simultaneous estimation and control algorithm given by problem \eqref{eq_mhe mpc opt problem} will be robustly stable.}
\end{theorem}

\noindent \textbf{Proof.} See Appendix \ref{Proof_T1}. {\hfill$\square$} \vspace{0.1cm}

\textcolor{blue}{Theorem \ref{theorem: forward window} links the robust stability of the \smhec{} with the length of the estimation and control windows without requiring the MPC's cost-to-go to be a CLF. It is worth noting that robust stability is guaranteed under mild practical assumptions. It can be stated as follows: \textit{given an i-IOSS system  with control actions large enough to mitigate the deviation due to process disturbances, then, a \smhec{} with estimation and control windows long enough will be robust and stable.} Theorem \ref{theorem: forward window} establishes how long the backward and forward windows should be.}

\textcolor{blue}{As expected, {a larger} uncertainty $e_{\max}$ on the initial state of the system {involves} a longer estimation window. However, MHE technique would enable the possibility of shortening the backward window after the initial uncertainty is reduced. {To the best of the authors' knowledge,} this topic deserves further research since it could have a deep practical meaning.
In the same way, the length of the control window increases as the requirement on the cost-to-go of being a CLF is relaxed. Moreover, when the amplitude of control actions are comparable with those of the disturbances, the length of the forward window needs to be enlarged too.
}

% Examples____________________________________________________
\section{Examples} \label{section:examples}

In this section, we discuss two examples to illustrate the results presented previously and the performance of the {proposed} framework{is analysed}. The first example applies the ideas introduced in previous sections to a nonlinear scalar system. The emphasis is placed in the effect of constraints and disturbances on closed-loop stability and performance. The second example discusses the simulations results for a van der Pol oscillator using the \smhec{}. The discussion is focused on the effect of $N_e$ and $N_c$ on the performance and computational time.

\subsection{Example 1}      \label{example: 1}

Let us consider the continuous--time nonlinear system 

\begin{equation}        \label{eq:system example 1}
    \begin{array}{rl}
        \Dot{x} =& a x^3_t + w_t + u_t, \quad a\in\mathbb{R}_{>0}\\
        y_t     =& x_t + v_t.
    \end{array}
\end{equation}

Firstly, we show its detectability, i.e., the existence of an estimate with a structure like equation \eqref{eq: definicion i-IOSS}. Let assume two arbitrary and feasible trajectories $x^{(1)}_t$ and $x^{(2)}_t$ such that $\Delta x \coloneqq x^{(1)}_t -x^{(2)}_t$ and $p_t \coloneqq \vert \Delta x \vert$; then $\Dot{p}_t$ can be written as follows

\begin{equation}
    \Dot{p}_t = \frac{\Delta x}{\vert \Delta x \vert}\left(\Dot{x}^{(1)}_t-\Dot{x}^{(2)}_t\right).
\end{equation}

Assuming a \emph{LTV} control law $u_t=-K_t x_t$, we obtain

\begin{equation}
    \Dot{p}_t = \frac{\Delta x}{\vert \Delta x \vert}\left(a\Delta x\left(x^{(1)^2}_t+x^{(1)}_t x^{(2)}_t+x^{(2)^2}_t\right)-K_t \Delta x+\Delta w_t\right),
\end{equation}
which is upper bounded by

\begin{equation}
    \Dot{p}_t \leq -K_t \, p_t + a \, g\vert \Delta h_t \vert  +\vert\Delta w_t \vert,
\end{equation}
where

\begin{equation}
    g\coloneqq h^2\left(x^{(1)}_0\right)+h\left(x^{(1)}_0\right) h\left(x^{(2)}_0\right) + h^2\left(x^{(2)}_0\right).   
\end{equation}
Solving $p_t$ for initial condition $p_0 = \vert x^{(1)}_0-x^{(2)}_0\vert$ we obtain

\begin{equation}   \label{eq:i-ioss bound example 1}
    \vert x^{(1)}_t - x^{(2)}_t \vert \leq \, \vert x^{(1)}_0 - x^{(2)}_0 \vert e^{-K_t t} + \frac{\Vert \Delta w_{0:t} \Vert}{K_t} + \frac{a g \Vert \Delta y_{0:t} \Vert}{K_t},
\end{equation}
it follows the fact that system \eqref{eq:system example 1} is \textit{i}-\emph{IOSS} (for all details, the reader can refer to appendix \ref{Deriv_p}). 

In the case of \emph{MHE--MPC} controllers with quadratic costs

\begin{equation}        \label{stag&term_costs}
 \begin{array}{rl}
  \ell_e \coloneqq \hat{w}^2_{j|k} Q_e + \hat{v}^2_{j|k} R_e, &
  \Gamma_{k - N_e} \coloneqq P^{-1}_{k - N_e} \chi^2, \\
   \ell_c \coloneqq \hat{x}^2_{j|k} Q_c + \hat{u}^2_{j|k} R_c, &
   \Upsilon_{k+N_c} \coloneqq S_{c} \Xi^2,
  \end{array}
\end{equation}
the bound \eqref{eq:i-ioss bound example 1} can be written as follows

\begin{equation}    \label{eq:estimation error bound example 1}
  \begin{array}{rl}
    \vert x_k-\hat{x}_{k\vert k} \vert \leq&\, \vert x_0-\bar{x}_0 \vert \left(\frac{\theta\,\mathbb{N}_e}{2N_e}\right)^i\left(2+\frac{\left(P^{-1}_{k-N_e}R_e\right)^{1/2}+\left(P^{-1}_{k-N_e}Q_e\right)^{1/2} \,a \, g}{\left(Q_eR_e\right)^{1/2}K^{c}_{1\vert k}}\right)
    + 2\left(1+\mu\right)\Vert \boldsymbol{w} \Vert \left( \frac{2}{K^{c}_{1\vert k}} + \frac{\left(Q_e \, R_e\right)^{1/2} K^{c}_{1\vert k} +\left(P^{-1}_{k-N_e} \, Q_e\right)^{1/2} \, a \, g}{\left(P^{-1}_{k-N_e} \, Q_e \right)^{1/2} K^{c}_{1\vert k}} \right) \vspace{2mm} \\
    &
    + 2\left(1 + \mu\right)\Vert\boldsymbol{v} \Vert \left(\frac{2 \, g }{K^{c}_{1\vert k}}+\frac{\left(R_e \, Q_e\right)^{1/2} K^{c}_{1\vert k} + \left(P^{-1}_{k-N_e} \, R_e\right)^{1/2}}{\left(P^{-1}_{k-N_e}Q_e\right)^{1/2} K^{c}_{1\vert k}}\right),
  \end{array}
\end{equation}
with $\mathbb{N}_e$ given by

\begin{equation}       \label{eq:Ne example 1}
    \mathbb{N}_e = \ceil*{4\left( 2 + \frac{\left(P^{-1}_{k-N_e} \, R_e\right)^{1/2}+\left( P^{-1}_{k-N_e} \, Q_e \right)^{1/2} a \, g}{\left( Q_e \, R_e \right)^{1/2}K^{c}_{1\vert k}} \right)^2},
\end{equation}
and $K^{c}_{1\vert k}$ is the equivalent controller gain resulting from applying $\hat{u}_{1|k}$.

Equations \eqref{eq:estimation error bound example 1} and \eqref{eq:Ne example 1} show the influence of the controller on the state estimation. Larger controller gains improve estimation error and shorten the convergence time. However, controller gains are bounded by stability conditions and input constraints, which are limiting factors in this potential improvement. This example highlights the relevance of simultaneously solving the estimation and control problems, or at least to take into account the solution of control problem on the estimation one. Since \emph{MPC} gains $K^{c}_{1 \vert k}$ are time-varying because they are recomputed at every sampling time, a conservative approach can employ its lowest value.

In order to compare the performances of independent and simultaneous \emph{MHE--MPC}, both output--feedback controllers have the same parameters

\begin{equation}        \label{Parameter_MPC}
   P_0=10^5,\, Q_e=15,\, R_e=10^3,\, Q_c=5,\, R_c=5,\, S_c=Q_c,\, \mu=0.05,
\end{equation}
with constraints sets

\begin{equation}        \label{Constr_S01}
  \mathcal{X}\coloneqq\left\{x:\vert x \vert \leq 0.8 \right\} 
  \text{ and }
  \mathcal{U}\coloneqq\left\{u:\vert u \vert \leq 2.5 \right\},    
\end{equation}
$w_t \sim \textnormal{\textit{U}}\left(0,\,0.01\right)$, $v_t \sim \mathcal{N}\left(0,\,0.02^2\right)$, $a=1$, $g=3x_{0 \, \max}^3$ and $\varphi=0.5$ such that both controllers implement the same optimisation criterion.

The control problems of both controllers are configured without terminal constraints. The process disturbance is not taken into account to compute $\hat{u}_{j|k},$ but it will be considered in the computation of $N_c$. It can be computed directly from equation \eqref{eq:stabilizinc lenght contrl window} once the values of $\delta$, $L$ and $\Delta^w_c$ had been established. Another approach, employed in this example, consists of computing $\omega$ through simulations. In this example, we set the initial condition that maximizes the controller costs and then computes the values of $L$ and $\omega$. The process is repeated until the maximal value of $N_c$ {is reached}.

\begin{figure}[b]
    \centering
    \begin{subfigure}[b]{0.45\textwidth}
        \centering
        \includegraphics[width=\textwidth]{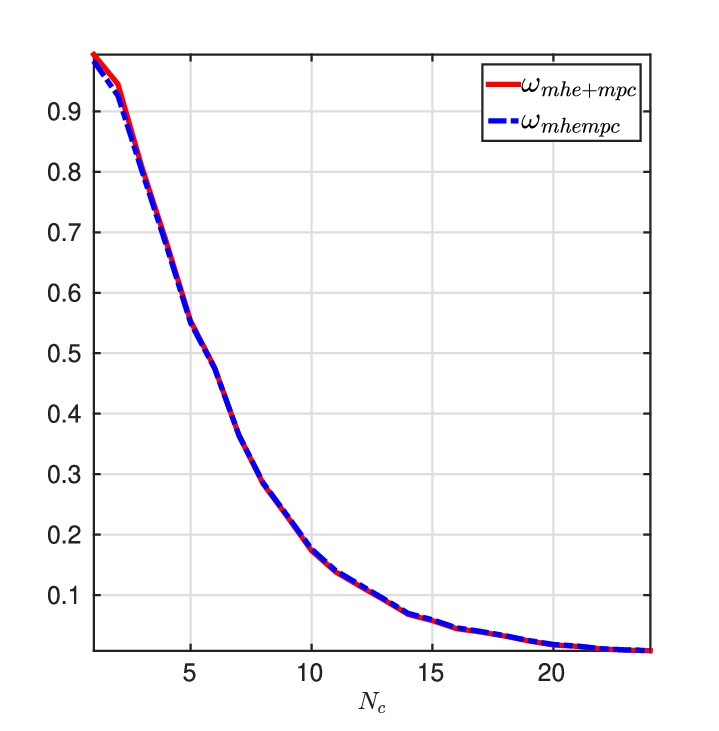}
        \caption{}
        \label{figure:example1 omega-a}
    \end{subfigure}
    \hfill
    \begin{subfigure}[b]{0.45\textwidth}
        \centering
        \includegraphics[width=\textwidth]{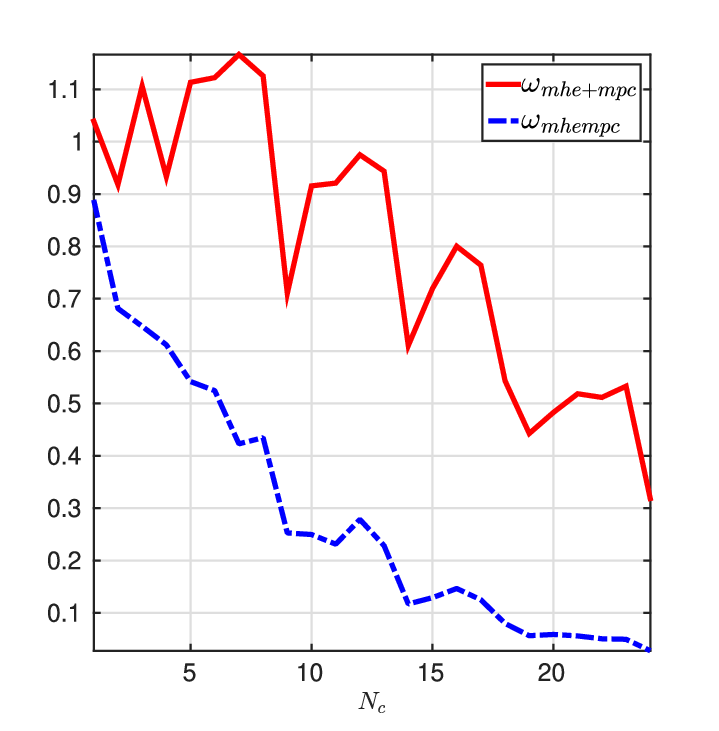}
        \caption{}
        \label{figure:example1 omega-b}
    \end{subfigure}
  \caption{Evolution of ${\omega}(N_c)$ for $\delta=1$ and $\Delta^w_c=10^{-1}$: (\ref{figure:example1 omega-a}) with constraints set \eqref{Constr_S01}, and (\ref{figure:example1 omega-b}) with constraint set \eqref{Constr_S02}.}
  \label{figure:example1 omega}
\end{figure}

\begin{figure}[t]
    \centering
    \begin{subfigure}{0.45\textwidth}
        \centering
        \includegraphics[width=\textwidth]{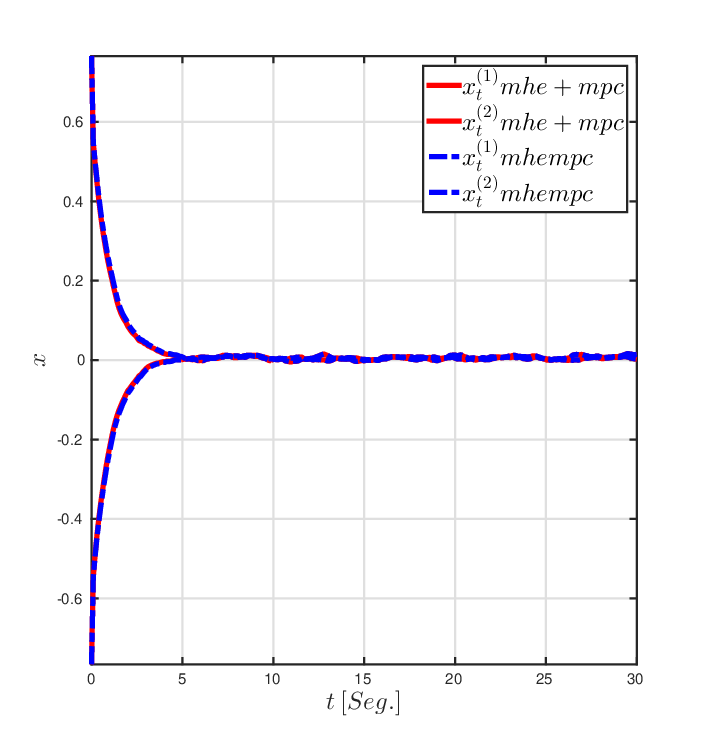}
        \caption{}
        \label{figure:i-ioss bound example 1-a}
    \end{subfigure}
    \hfill
    \begin{subfigure}{0.45\textwidth}
        \centering
        \includegraphics[width=\textwidth]{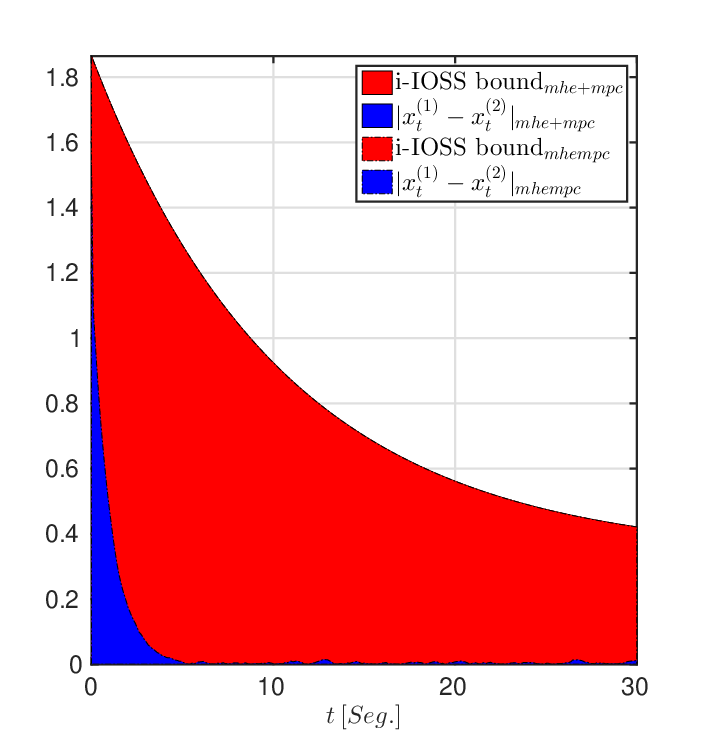}
        \caption{}
        \label{figure:i-ioss bound example 1-b}
    \end{subfigure}
    \caption{Evolution of system output for different initial conditions (\ref{figure:i-ioss bound example 1-a}) and difference between trajectories and \textit{i}-\emph{IOSS} bound (\ref{figure:i-ioss bound example 1-b}).}
    \label{figure:i-ioss bound example 1}
\end{figure}

\begin{figure}[b]
    \centering
    \begin{subfigure}{0.42\textwidth}
        \centering
        \includegraphics[width=\textwidth]{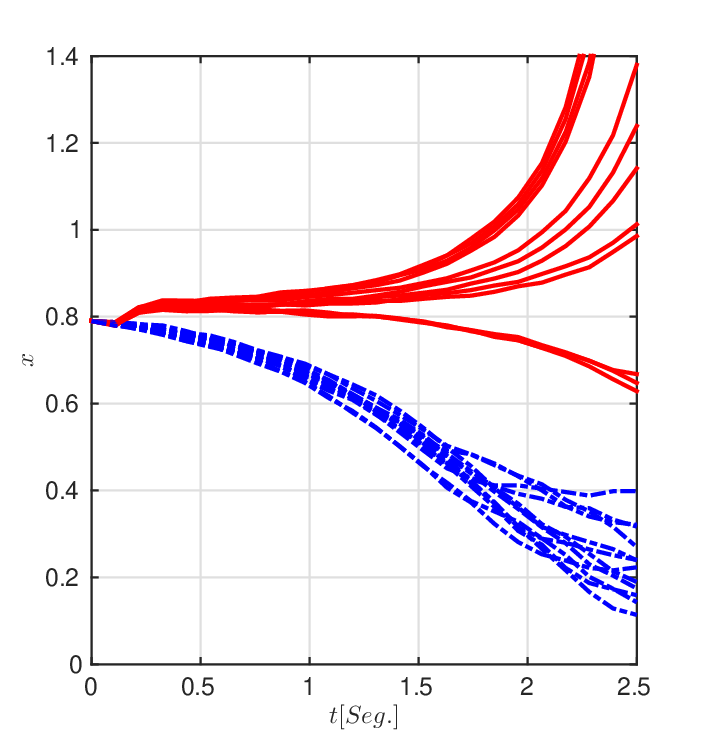}
        \caption{}
        \label{figure:larger control horizons-a}
    \end{subfigure}
    \hfill
    \begin{subfigure}{0.42\textwidth}
        \centering
        \includegraphics[width=\textwidth]{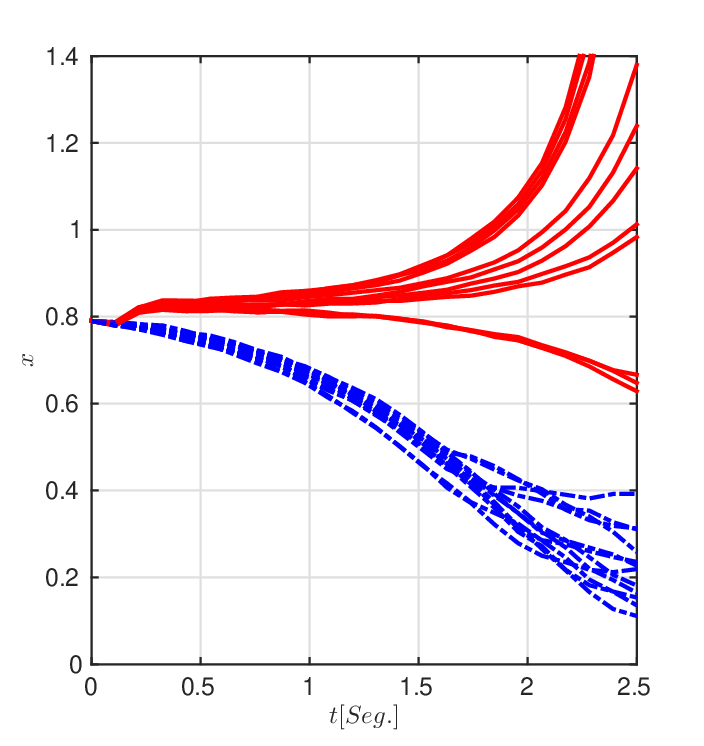}
        \caption{}
        \label{figure:larger control horizons-b}
    \end{subfigure}
    \caption{Evolution of system output for $N_c=20$ (\ref{figure:larger control horizons-a}) and $N_c=70$ (\ref{figure:larger control horizons-b}), with $N_e=30$ for independent \emph{MHE} and \emph{MPC} (continuous red line) and simultaneous \emph{MHE--MPC} (blue {dashed} line).}
  \label{figure:larger control horizons}
\end{figure}

Figure \ref{figure:example1 omega} shows the computed values of ${\omega}$ {as} a function of $N_c$ ($\omega(N_c)$) for the same $\Delta^w_c$ and different set of constraints and distributions for process and measurement noises. In this figure the effect of constraints on ${\omega}(N_c)$ can be seen: They increase ${\omega}(N_c)$, for the same $N_c$, depending how the controller is implemented. This change is smaller for the simultaneous \emph{MHE--MPC} approach than the independent one. When constraints are no relevant (constraints set \eqref{Constr_S01}), both controllers have similar values (see Figure \ref{figure:example1 omega-a}), and the control problem of both can use the same $N_c$. However when constraints \eqref{Constr_S02} are considered, the way of solving the estimation and control problems has a direct effect on ${\omega}(N_c)$ (see Figure \ref{figure:example1 omega-b}), and the control problem of both controllers must use different $N_c$ in order to ensure stability, affecting the computational requirements of the implementation. Since we are using constraints set \eqref{Constr_S01} we choose $N_c=10$ for both controllers (Figure \ref{figure:example1 omega-a}). Finally, the minimum estimation horizon $\mathbb{N}_e$ is computed from \eqref{eq:minimo N} using the parameters listed in \eqref{Parameter_MPC}, leading to $\mathbb{N}_e=27$ for both controllers. We choose $N_e=30$ for both controllers.

Figure \ref{figure:i-ioss bound example 1} shows the system responses and the corresponding \textit{i}-\emph{IOSS} bound for the regulation problem. Figure \ref{figure:i-ioss bound example 1-a} shows two trajectories generated by both controllers from different initial condition ($x_0^{(1)}=0.766$ and $x_0^{(2)}=-0.766$) with the same prior guess ($\bar{x}_0=-2.5$). Figure \ref{figure:i-ioss bound example 1-b} shows the difference between the trajectories and its \textit{i}-\emph{IOSS} bound, for the minimum controller gain along the simulation ($K^{c}_{1|k}=0.7326$). {From} this figure {it can be seen} the decreasing behaviour of the estimation error bound, as expected from equation \eqref{eq: bound estimation error} for the general case and \eqref{eq:estimation error bound example 1} for this particular example. Despite the {fact that the} value of $\mu$ {is small} ($\mu=0.05$), the bound \eqref{eq:estimation error bound example 1} is quite conservative. In these figures, we can also see that both controllers provide a similar response, since constraints and disturbances have not relevant effect on the system behaviour, and therefore the separation principle can be applied.

Now let us compare the performance in a more challenging setup. In the following, we will assume the next constraints set

\begin{equation}        \label{Constr_S02}
  \mathcal{U}\coloneqq\left\{u:\vert u \vert \leq 0.6 \right\},
  \mathcal{W}\coloneqq\left\{w:\vert w \vert \leq 0.4 \right\}
  \text{ and }
  \mathcal{V}\coloneqq\left\{v:\vert v \vert \leq 0.8 \right\}.
\end{equation}

The controls $\hat{u}_{j|k}$ have been tightened and the estimates $\hat{w}$ and $\hat{v}$ have been constrained to the sets $\mathcal{W}$ and $\mathcal{V}$, respectively. Disturbances $w_t$ and $v_t$ are now given by $w_t \sim \textnormal{\textit{U}}\left(0,\,0.1\right)$ and $v_t \sim \mathcal{N}\left(0,\,0.2^2\right)$, respectively.

Under this new operational conditions $\mathbb{N}_e$ is recomputed, obtaining $\mathbb{N}_e=98$ for the independent \emph{MHE} and \emph{MPC}, and $\mathbb{N}_e=52$ for the simultaneous \emph{MHE--MPC}. This is the effect of constraints set \eqref{Constr_S02} on the estimator parameters, while the effect on the controller is shown in Figure \ref{figure:example1 omega-b}. This figure shows that the independent \emph{MHE} and \emph{MPC} approach is more sensitive to disturbances, requiring conservative values of $N_c$ to guarantee the closed--loop stability.

Finally, Figure \ref{figure:larger control horizons} shows the system responses for regulation problem for different realisations of $w_t$ and $v_t$ and different $N_c$, for $N_e=30$. The independent \emph{MHE} and \emph{MPC} strategy fails to regulate the system states for some noise realisations, even though it regulates few of them. On the other hand, the simultaneous \emph{MHE--MPC} controller manages to regulate the system states for all noise realisations. This problem is caused by the failure of the independent \emph{MHE} and \emph{MPC} to satisfy Assumption \ref{assumption: feasibility condition}. In fact, its design procedure applies the \textit{separation principle}, which entails the automatic satisfaction of Assumption \ref{assumption: feasibility condition} and it does not include the constraints information in the selection of $N_e$ and $N_c$, while the simultaneous \emph{MHE--MPC} controller does.

\subsection{Example 2}      \label{example: 2}
Let us consider the van der Pol oscillator whose dynamic is described by

\begin{equation}        \label{Osc_vdP}
 \begin{array}{rl}
  \Dot{x}_t =& \left[\begin{array}{c}
  \epsilon\left(1-x_{2,t}^2 \right)x_{1,t}-2x_{2,t}+u_t+w_{1,t} \\
              2x_{1,t}+w_{2,t}\end{array}\right] \quad \epsilon\in\mathbb{R}_{\geq 0},\vspace{1.5mm} \\
        y_t =& \frac{1}{2}\left(x_{1,t}+x_{2,t}\right)+v_{1,t}.
 \end{array}
\end{equation}
It is known to be \textit{i}-\emph{IOSS}, and a proof of this property can be made using the averaging lemma \citep{pogromsky2015stability}. In this example we will focus the analysis on the system performance under different set of parameters. The independent and simultaneous \emph{MHE--MPC} controllers have the same parameters to allow a direct comparison of their performances. All the stage costs have a quadratic structure (equation \eqref{stag&term_costs}) and their parameters are

\begin{equation}        \label{Parameter_MPC02}
   P_0=10^5,\, Q_e=\begin{bmatrix} 50 & 0\\0 & 50 \end{bmatrix},\ R_e=150,\, Q_c=\begin{bmatrix} 200 & 0\\0 & 200 \end{bmatrix},\,S=Q_c,\, R_u=10^{-2},
\end{equation}

\noindent with constraints sets given by

\begin{equation}        \label{Const_O}
   \mathcal{X} \coloneqq \left\{x:\vert x_1 \vert\leq 5, \vert x_2\vert\leq 5\right\},\, \mathcal{U} \coloneqq \left\{u:\vert u \vert \leq 5, \vert \Delta u_k \vert\leq 2 \right\},
\end{equation}

\noindent $w_t \sim \textnormal{\textit{U}}\left(0,0.25\right)$ and $v_t \sim \textnormal{\textit{U}}\left(0,0.025\right)$, instead of zero mean normal distribution, as it is common in the literature.

{\color{blue}
The minimum estimation ($\mathbb{N}_e$) and control ($\mathbb{N}_c$) horizons were computed using equations (\ref{eq:stabilizinc lenght contrl window}) and parameters given by (\ref{Parameter_MPC02}), leading to $\mathbb{N}_e = 10$ and $\mathbb{N}_c = 15$. However, different values of $N_e$ and $N_c$ were chosen 

\begin{equation} \label{eq:ne_nc_example2}
   N_e \coloneqq \{2,5,10,20\},\, N_c \coloneqq \{5,10,35\}.
\end{equation}
to analyse their effect on the closed-loop performance.}

Since the difference between $N_e$ and $N_c$ can lead to unbalanced cost functions (emphasizing the control cost over the estimation one), which can deteriorate the overall closed-loop performance. To avoid this problem, $\varphi$ is used to improve the closed-loop performance. It takes the following values $\varphi \coloneqq \{0.95,0.95,0.85,0.65\}$ for the corresponding $N_e$ value.

\begin{figure}[thb]
    \centering
    \begin{subfigure}{0.40\textwidth}
        \centering
        \includegraphics[width=0.80\textwidth]{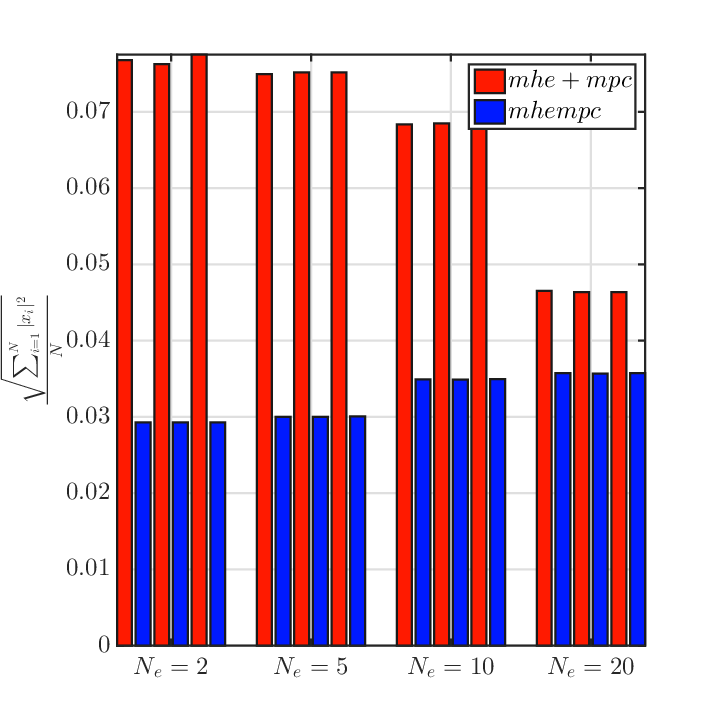}
        \caption{}
        \label{figure:example 2, epsilon=1-3-a}
    \end{subfigure}
    \begin{subfigure}{0.40\textwidth}
        \centering
        \includegraphics[width=0.80\textwidth]{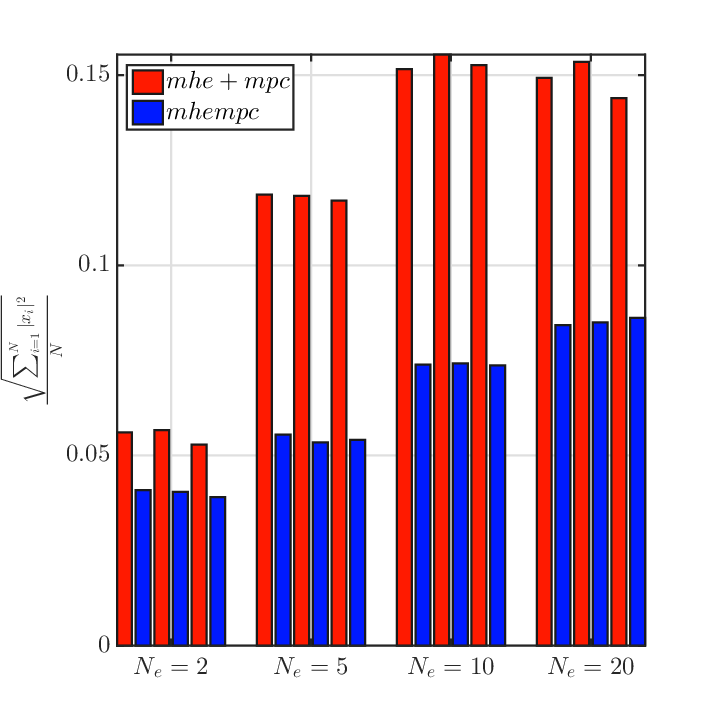}
        \caption{}
        \label{figure:example 2, epsilon=1-3-b}
    \end{subfigure}
  \caption{\emph{MSE} of $100$ simulations for different values of $N_e$, $N_c$ and $\epsilon$}
  \label{figure:example 2, epsilon=1-3}
\end{figure}

\begin{figure}[b]
    \centering
    \begin{subfigure}[t]{0.40\textwidth}
        \centering
        \includegraphics[width=\textwidth,height=145pt]{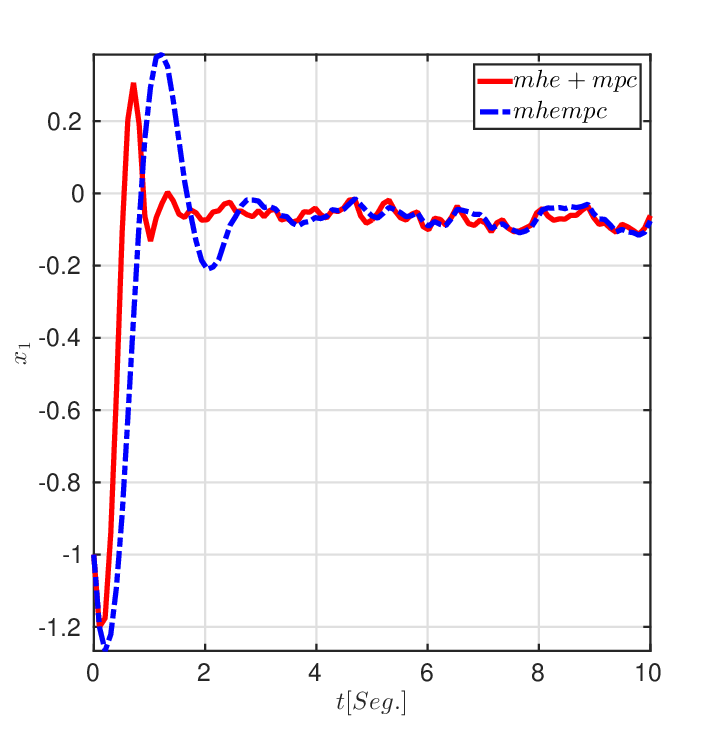}
        \caption{}
        \label{figure:example 2, epsilon=1 x1 and x2-a}
    \end{subfigure}
    \hfill
    \begin{subfigure}[t]{0.40\textwidth}
        \centering
        \includegraphics[width=\textwidth,height=145pt]{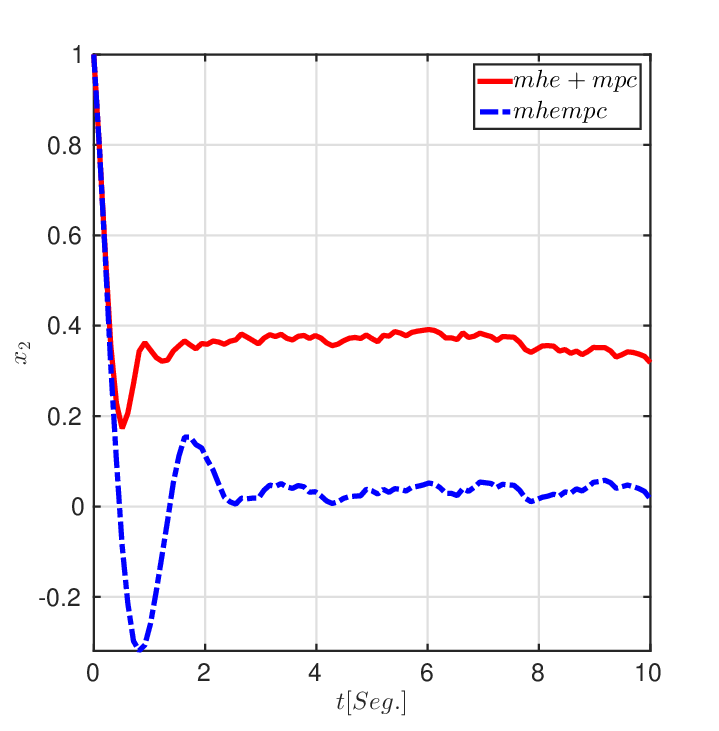} 
        \caption{}
        \label{figure:example 2, epsilon=1 x1 and x2-b}
    \end{subfigure}\\
    \begin{subfigure}[b]{0.40\textwidth}
        \centering
        \includegraphics[width=\textwidth,height=145pt]{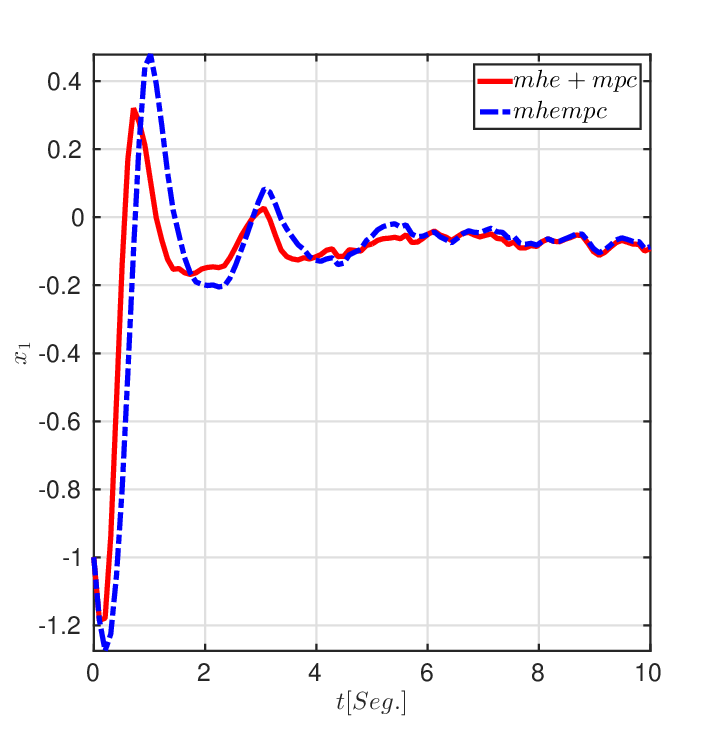}
        \caption{}
        \label{figure:example 2, epsilon=1 x1 and x2-c}
    \end{subfigure}
    \hfill
    \begin{subfigure}[b]{0.40\textwidth}
        \centering
        \includegraphics[width=\textwidth,height=145pt]{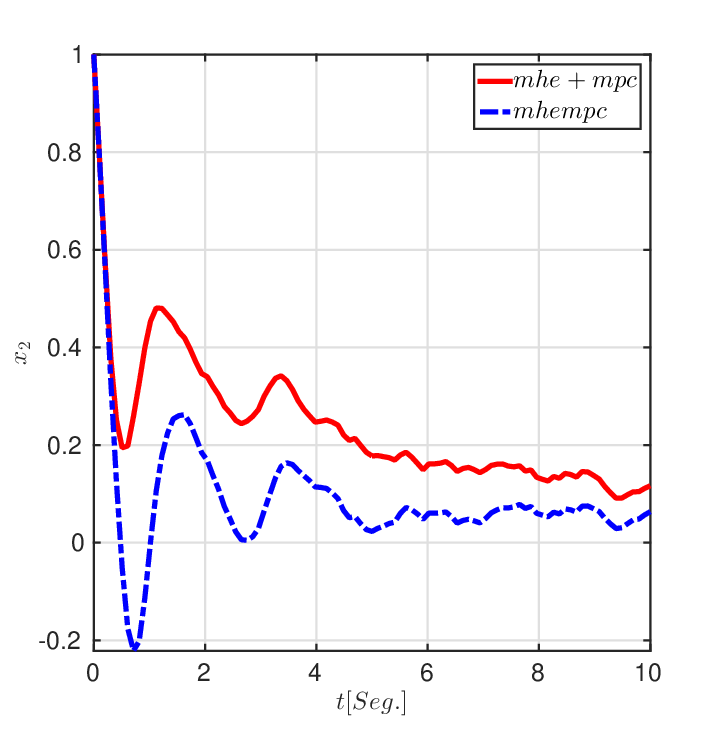} 
        \caption[l]{}
        \label{figure:example 2, epsilon=1 x1 and x2-d}
    \end{subfigure}
    \caption{Two realizations of $x_1$ and $x_2$ for $\epsilon=0.1$, $N_e=2$ (\ref{figure:example 2, epsilon=1 x1 and x2-a} - \ref{figure:example 2, epsilon=1 x1 and x2-b}), $N_e=20$ (\ref{figure:example 2, epsilon=1 x1 and x2-c} - \ref{figure:example 2, epsilon=1 x1 and x2-d}) and $N_c=35$.}
    \label{figure:example 2, epsilon=1 x1 and x2}
\end{figure}

\begin{figure}[thb]
  \centering
  \begin{tabular}{c}
    {\includegraphics[width=0.6\textwidth]{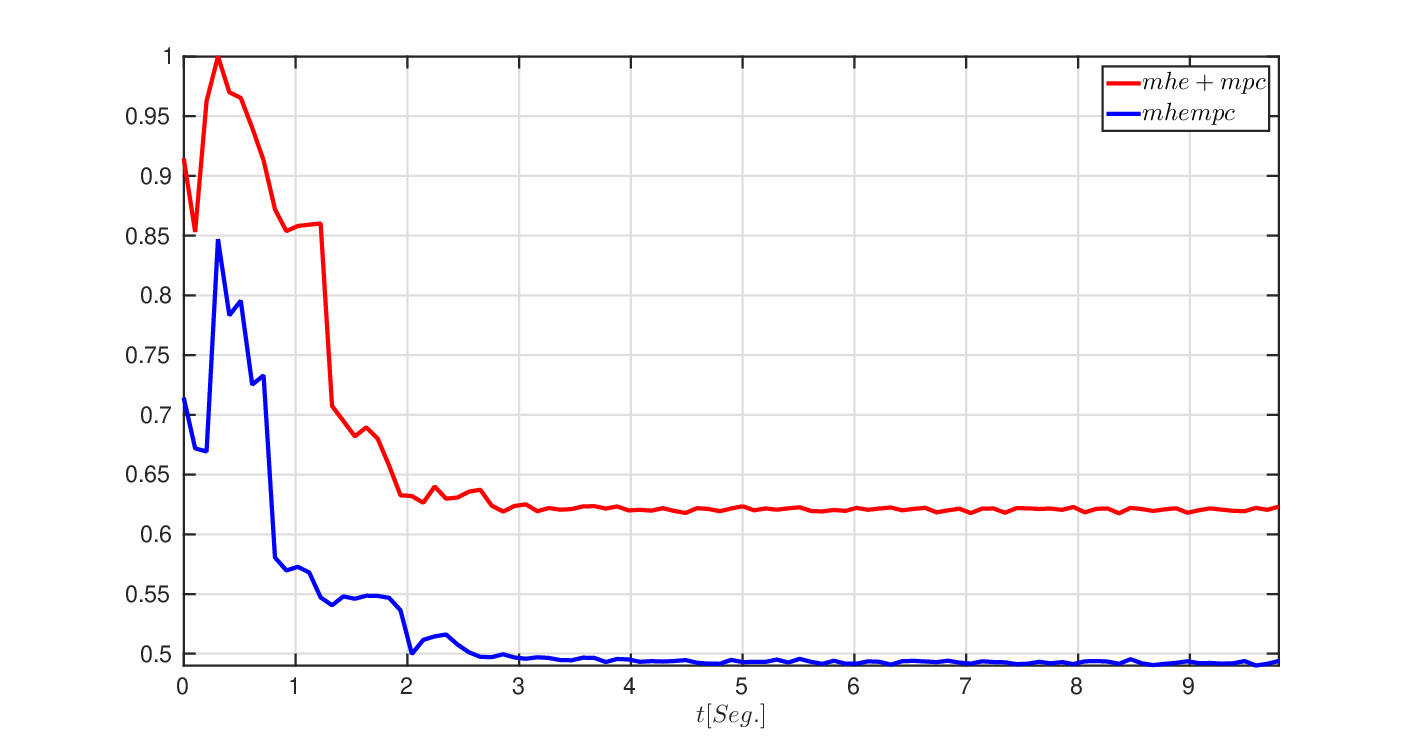}}
  \end{tabular}
  \caption{Average execution time in seconds over 100 trials for $N_e=N_c=10$.}
  \label{figure:execution times}
\end{figure}

Figures \ref{figure:example 2, epsilon=1-3} summarize the mean square error (\emph{MSE}) obtained by both controllers along $100$ simulations for $\epsilon = 0.1$ and $\epsilon = 3$ respectively. These figures show the superior performance of the simultaneous \emph{MHE--MPC} for any combination of $N_e-N_c$ and scenario. In general, there are no meaningful changes of \emph{MSE} with $N_c$, however closed-loop performance varies with $N_e$. Figure \ref{figure:example 2, epsilon=1-3-a} shows the results for $\epsilon=0.1$. In this case the independent \emph{MHE} and \emph{MPC} performance improves with $N_e$, while the simultaneous \emph{MHE--MPC} ones remains similar (a deviation lower than $8\%$ from the average) for any combination of $N_e-N_c$. For this value of $\epsilon$, the system \eqref{Osc_vdP} behaves like a harmonic oscillator, therefore the closed-loop performance depends on the estimation error (see Figure \ref{figure:example 2, epsilon=1 x1 and x2}), which decreases for larger values of $N_e$. 
Figure \ref{figure:example 2, epsilon=1-3-b} shows the results for $\epsilon=3$. In this condition, the performance of both controllers deteriorates with $N_e$, because for this value of $\epsilon$ the system \eqref{Osc_vdP} behaves like a non-linear dampened oscillator and the state estimates take longer to converge to the estimation invariant set (see Figures \ref{figure:example 2, epsilon=1 x1 and x2-a} and \ref{figure:example 2, epsilon=1 x1 and x2-b}).

Figures \ref{figure:example 2, epsilon=1 x1 and x2} show the simulations resulting from two noise realizations for $N_c=35$, $N_e=2$ and $N_e=20$, respectively. They show that the \smhec{} manages to regulate both states and it achieves a better performance than the independent one. While Figures \ref{figure:example 2, epsilon=1 x1 and x2-a} and \ref{figure:example 2, epsilon=1 x1 and x2-c} show that independent  \emph{MHE} and \emph{MPC} achieves a better performance than the simultaneous one for state $x_1$, Figures \ref{figure:example 2, epsilon=1 x1 and x2-b} and \ref{figure:example 2, epsilon=1 x1 and x2-d} show how it fails to regulate state $x_2$ for short estimation horizons. Under this condition, $x_2$ has an offset that it is not compensated by the controller. Only large values of $N_e$ allow the independent \emph{MHE} and \emph{MPC} to regulate $x_2$ (Figure \ref{figure:example 2, epsilon=1 x1 and x2-d}). On the other hand, the \smhec{} regulates both states and it takes shorter times than the independent one to regulate both states.

The computational burden of the simultaneous \emph{MHE--MPC} is lower than the independent one, as can be seen in Figure \ref{figure:execution times}. The execution times were averaged over $100$ trials. The lower time, in the beginning, is due to the backward window corresponding to the estimation has not achieved yet its full length.

% Conclusions _________________________________________________________
\section{Conclusions} \label{section:conclusion}
We presented an output-feedback approach for nonlinear systems subject to bounded disturbances using \emph{MHE--MPC}. The proposed approach combines the state estimation and control problems into a single optimisation, which is solved at each sampling time. Theorem  \ref{theorem: forward window} states the necessary conditions to guaranty the feasibility and stability of the optimisation problem, and therefore the boundedness of system states, as a function of the windows lengths $N_e$ and $N_c$. This result requires the compatibility between the robust estimated and controllable sets (Assumption \ref{assumption: feasibility condition}) and the existence of a relaxed closed--loop Lyapunov function for the disturbed system (Assumption \ref{relaxed contraction of Vf}). These conditions imply forward ($N_c$) and backward ($N_e$) horizons to find state estimates and control actions that are consistent with the system dynamics, constraints and disturbances.

\textcolor{blue}{Even though the stability and performance of the simultaneous approach are seen to outperform the \textit{MHE} and \textit{MPC} solved independently, the simultaneous method has shown a higher sensibility to the parameters of the backward and forward windows, requiring a more detailed and precise tuning process. The stage-cost matrices on both windows need to be chosen carefully. The parameter $\varphi$ can help in the tuning procedure. However, a high level of expertise is required in order to succeed with the adjustment of matrices.}

Future work may involve the design of the forward window with properties that allow the improvement of the estimation process and the design of an adaptive law to compute $\varphi$ such that the estimation and control problems keep balanced and the overall system performance and numerical properties are improved.

\section*{Acknowledgments}
The authors wish to thank the Consejo Nacional de Investigaciones Científicas y Técnicas (CONICET) from Argentina, for their support. 

\section*{Conflict of Interest Statement}
The authors declare no potential conflict of interests.

\section*{Data Availability Statement}
The data that support the findings of this study are available from the corresponding author upon reasonable request.

\appendix

\section{ Proof Theorem 1}   \label{Proof_T1}
In the following we will analyse the stability of the simultaneous \emph{MHE--MPC} algorithm by means of the difference in costs at two consecutive sampling time 

\begin{equation}
 \Delta \Psi = \Psi_{\textnormal{\fontsize{6pt}{8pt}\selectfont\textit{EC,k+1,$N_e+N_c$}}}-\Psi_{\textnormal{\fontsize{6pt}{8pt}\selectfont\textit{EC,k,$N_e+N_c$}}}.   
\end{equation} 

Evaluating $\Psi_{\textnormal{\fontsize{6pt}{8pt}\selectfont\textit{EC,k+1,$N_e+N_c$}}}$ with the tail of the solution computed at time $k$, with $\hat{u}_{k+N_c}=0$ and $\hat{x}_{k+N_c+1}=f(\Xi,\hat{u}_{k+N_c})$, we obtain

\begin{equation}        \label{eq:costs comparison}
  \begin{array}{rl}
    \Delta \Psi =& \Gamma_{k-N_e+1}\left(\chi_{k-N_e+1} \right) + \displaystyle \sum_{j=k-N_e+1}^{k} \ell_{w_e} \left(\hat{w}_{j\vert k+1}\right) + \displaystyle \sum_{j=k-N_e+1}^{k+1} \ell_{v_e}\left(\hat{v}_{j\vert k+1} \right) \\ 
    & \displaystyle + \sum_{j=k+1}^{k+N_c}\left(\ell_c \left(\Hat{x}_{j\vert k+1}, \hat{u}_{j\vert k+1} \right) - \ell_{w_c}\left( \hat{w}_{j\vert k+1} \right)\right) + 
    \Upsilon_{k+N_c+1}\left(f(\Xi,\hat{u}_{k+N_c}) \right) \\
    & - \left(\Gamma_{k-N_e}\left(\chi \right)+\displaystyle \sum_{j=k-N_e}^{k-1} \ell_{w_e}
    \left(\hat{w}_{j\vert k}\right) + \sum_{j=k-N_e}^{k}\ell_{v_e}\left( \hat{v}_{j\vert k} \right)\right. \\ 
    &\left. \displaystyle + \sum_{j=k}^{k+N_c-1}\left(\ell_c \left(\Hat{x}_{j\vert k}, \hat{u}_{j\vert k} \right) - \ell_{w_c}\left( \hat{w}_{j\vert k} \right)\right) + \Upsilon_{k+N_c}\left(\Xi\right)\right).
  \end{array}
\end{equation}

Since $\chi_{k-N_e+1}= \hat{x}_{k-N_e+1\vert k+1}-\bar{x}_{k-N_e+1}$ and

\begin{equation}
 \begin{array}{rl}
  \bar{x}_{k-N_e+1}=&\hat{x}_{k-N_e+1\vert k}, \\
  \hat{x}_{k-N_e+1\vert k+1}=&\hat{x}_{k-N_e+1\vert k},
 \end{array}
\end{equation}
then $\Gamma_{k-N_e+1}(\chi_{k-N_e+1})=0$. Using inequality \eqref{relaxed contraction of Vf} and Assumption 5, $\Delta \Psi$ can be rewritten as follows

\begin{equation}        \label{eq:costs comparison 1}
  \begin{array}{rl}
    \Delta \Psi \leq& -\ell_c\left( \hat{x}_{k\vert k}, \hat{u}_{k\vert k} \right)\left( 1 - \delta \left( \displaystyle \frac{\Upsilon_{k+N_c}\left(\Xi\right)}{\ell_c\left( \hat{x}_{k\vert k}, \hat{u}_{k\vert k} \right)} +  
    \frac{1}{\delta} \frac{\ell_{w_c}\left( \hat{w}_{k\vert k} \right)}{\ell_c\left( \hat{x}_{k\vert k}, \hat{u}_{k\vert k} \right)}
    \right) \right) \vspace{0.15cm} \\
    & -\Gamma_{k-N_e}\left(\chi\right) + \ell_{w_e}\left(\hat{w}_{k\vert k+1}\right) - \ell_{w_e}\left(\hat{w}_{k-N_e\vert k}\right)-\ell_{v_e}\left(\hat{v}_{k-N_e\vert k}\right),\\
    \leq& -\ell_c\left( \hat{x}_{k\vert k}, \hat{u}_{k\vert k} \right)\left( 1 - \delta \left( \displaystyle \frac{\Upsilon_{k+N_c}\left(\Xi\right)}{\ell_c\left( \hat{x}_{k\vert k}, \hat{u}_{k\vert k} \right)} + 
    \frac{1}{\delta}\, \Delta^w_c 
    \right) \right) \vspace{0.15cm} \\
    & -\Gamma_{k-N_e}\left(\chi\right) + \ell_{w_e}\left( \hat{w}_{k\vert k}\right) - \ell_e\left( \hat{w}_{k-N_e\vert k}, \hat{v}_{k-N_e\vert k} \right).
  \end{array}
\end{equation}
for $\delta \in \mathbb{R}_{> 0}$. Defining functions $\omega$ and $\pi_E$ as follows

\begin{equation}    \label{eq: diff costs}
 \begin{array}{rl}
  \omega \coloneqq& \displaystyle \frac{\Upsilon_{k+N_c}\left(\Xi\right)}{\ell_c\left( \hat{x}_{k\vert k}, \hat{u}_{k\vert k} \right)} +     \displaystyle \frac{1}{\delta} 
   \, \Delta^w_c,\vspace{0.1cm}  \\
  \pi_E \coloneqq & -\Gamma_{k-N_e}\left(\chi\right) + \ell_{w_e}\left( \hat{w}_{k\vert k}\right) - \ell_e\left( \hat{w}_{k-N_e\vert k}, \hat{v}_{k-N_e\vert k} \right),
 \end{array}
\end{equation}
the equation \eqref{eq:costs comparison 1} can be written in a compact way 

\begin{equation}        \label{eq:costs comparison 2}
  \Delta \Psi \leq -\ell_c\left( \hat{x}_{k\vert k}, \hat{u}_{k\vert k} \right)\left( 1 - \delta \omega \right) + \pi_E.
\end{equation}

The term $\omega$ quantifies the improvements in the control cost (through the ratio between the \textit{cost-to-go} $\Upsilon_{k+N_c}(\cdot)$ and the control stage cost $\ell_{c}(\cdot,\,\cdot)$ at time $k$) and the disturbance controllability (the ratio between the control stage costs $\ell_{w_c}(\cdot)$ and $\ell_c(\cdot,\,\cdot)$ at time $k$). 

The term $\pi_{E}$ quantifies the changes in the estimation cost by measuring the amount of information left behind the estimation window (the \textit{arrival--cost} $\Gamma_{k-N_e}(\cdot)$). Since $\hat{w}_{k\vert k}$ was computed within the control window (maximised), it tends to take larger values than $\hat{w}_{k-N_e\vert k}$ which was computed within the estimation window (minimised). Therefore, when state estimation is precise (i.e., $\Gamma_{k-N_e}(\chi)$ remains low), the term $\pi_E$ will tend to take positive values, whereas if a major correction is made on the initial condition $\hat{x}_{k-N_e\vert k}$ (i.e., $\Gamma_{k-N_e}(\chi)$ will take big values), the improvement in the estimated trajectory will lead a decreasing cost with sharper slope.

Since

\begin{equation}
\begin{array}{rl}
    \pi_E &= -\Gamma_{k-N_e}\left(\chi\right) + \ell_{w_e}\left( \hat{w}_{k\vert k}\right) - \ell_e\left( \hat{w}_{k-N_e\vert k}, \hat{v}_{k-N_e\vert k} \right),\\
    &\leq \ell_{w_e}\left( \hat{w}_{k\vert k} \right),\\
    &\leq \overline{\gamma}_{w_e}\left(\vert \hat{w}_{k\vert k}\vert\right),\\
    &\leq \overline{\gamma}_{w_e}\left(\Vert\boldsymbol{\hat{w}}\Vert\right),
\end{array}
\end{equation}

which can be written in term of $\mathcal{K}$ functions as follows \citep{deniz2019robust}

\begin{equation}
 \begin{array}{rl}
 \pi_{E} \leq \overline{\gamma}_w\left(\Vert \boldsymbol{\hat{w}} \Vert\right) \leq& \overline{\pi}_E \coloneqq \overline{\gamma}_w\left(\underline{\gamma}_w^{-1}\left( \frac{\overline{\gamma}_p(\chi)}{N_e}+\overline{\gamma}_w(\Vert \boldsymbol{w} \Vert)+\overline{\gamma}_v(\Vert \boldsymbol{v} \Vert) \right)\right),
 \end{array}
\end{equation}

Restating \eqref{eq:costs comparison 2} with $\overline{\pi}_E$, $\Delta \Psi$ can be posed as

\begin{align}
\label{eq:costs comparison 3}
    \Delta \Psi \leq& -\ell_c\left( \hat{x}_{k\vert k}, \hat{u}_{k\vert k} \right)\left( 1 - \delta \omega \right) + \overline{\pi}_E,
\end{align}

From the first term in the right hand side of \eqref{eq:costs comparison 3}, one can see that if 

\begin{equation}
\label{condition req for mpc stable}
   0 \leq \delta\omega < 1
\end{equation}
then, for large values of $\ell_c\left(\hat{x}_{k\vert k},\hat{u}_{k\vert k}\right)$ so that it becomes dominating in \eqref{eq:costs comparison 3}, the sequence of cost will present a contractive behaviour until $\ell_c\left(\hat{x}_{k\vert k},\hat{u}_{k\vert k}\right)(1-\delta\omega)$ reaches the value of $\overline{\pi}_E$, \textcolor{blue}{i.e., the decreasing behaviour in the costs is determined by the estimation error. Moreover, one is interested in guaranteeing that $0\leq \delta\omega<1$. In order to relate the quantity $\delta \omega$ with design parameter of the \smhec{} framework, let us recall that}

\begin{equation*}
    \begin{array}{rl}
         \omega =& \displaystyle \frac{\Upsilon_{k+N_c}\left(\Xi\right)}{\ell_c\left( \hat{x}_{k\vert k}, \hat{u}_{k\vert k} \right)} +     \displaystyle \frac{1}{\delta} \, \Delta^w_c,
    \end{array}
\end{equation*}
or

\begin{equation*}
    \begin{array}{rl}
         \delta\omega =& \displaystyle \frac{\delta\Upsilon_{k+N_c}\left(\Xi\right)}{\ell_c\left( \hat{x}_{k\vert k}, \hat{u}_{k\vert k} \right)} +     \displaystyle  \, \Delta^w_c,
    \end{array}
\end{equation*}

Therefore, the following needs to be guaranteed

\begin{equation}
    \begin{array}{rl}
         0\leq \displaystyle \frac{\delta\Upsilon_{k+N_c}\left(\Xi\right)}{\ell_c\left( \hat{x}_{k\vert k}, \hat{u}_{k\vert k} \right)} +     \displaystyle \Delta^w_c, < 1
    \end{array}
\end{equation}

The non negativity holds trivially since all quantities involved are positive. Therefore, the following is the inequality to fulfil 

\begin{align}
\label{eq:condition for stability 2}
    \displaystyle \frac{\Upsilon_{k+N_c}\left(\Xi\right)}{\ell_c\left( \hat{x}_{k\vert k}, \hat{u}_{k\vert k} \right)} <& \frac{1-\Delta^w_c}{\delta}
\end{align}

Since $0<\Delta^w_c<1$ by assumption \ref{assumption:condition for stability}, right hand side of inequality \eqref{eq:condition for stability 2} will be positive. The problem consists now in to find a value of $N_c$ such that \eqref{eq:condition for stability 2} be verified. In order to relate \eqref{eq:condition for stability 2} with $N_c$, let us note that

\begin{equation}        \label{**}
  \begin{array}{rl}
    \Psi_{C,k,N_c}  =& 
    \sum_{j=k}^{k+N_c-1}\left(\ell_c \left(\Hat{x}_{j\vert k}, \hat{u}_{j\vert k} \right) - \ell_{w_c}\left( \hat{w}_{j\vert k} \right)\right) + \Upsilon_{k+N_c}\left(\Xi\right),\vspace{2mm}\\
    =& \ell_c \left(\hat{x}_{k\vert k}, \hat{u}_{k\vert k} \right)\sum_{j=k}^{k+N_c-1}\frac{\left(\ell_c \left(\Hat{x}_{j\vert k}, \hat{u}_{j\vert k} \right) - \ell_{w_c}\left( \hat{w}_{j\vert k} \right)\right)}{\ell_c \left(\Hat{x}_{k\vert k}, \hat{u}_{k\vert k} \right)} + \frac{\Upsilon_{k+N_c}\left(\Xi\right)}{\ell_c \left(\Hat{x}_{k\vert k}, \hat{u}_{k\vert k} \right)},\vspace{2mm}\\
    \leq& \ell_c \left(\hat{x}_{k\vert k}, \hat{u}_{k\vert k} \right)\sum_{j=k}^{k+N_c-1}\frac{\left(\ell_c\hat{x}_{j\vert k}, \hat{u}_{j\vert k}\right)}{\ell_c \left(\Hat{x}_{k\vert k}, \hat{u}_{k\vert k} \right)} + \frac{\Upsilon_{k+N_c}\left(\Xi\right)}{\ell_c \left(\Hat{x}_{k\vert k}, \hat{u}_{k\vert k} \right)}.
  \end{array}
\end{equation}

{\color{blue}{

Then, the term $\frac{\Upsilon_{k+N_c}\left(\Xi\right)}{\ell_c \left(\Hat{x}_{k\vert k}, \hat{u}_{k\vert k} \right)}$ can be upper bounded as follows

\begin{align}   \label{eq: prod of cocientes}
     \frac{\Upsilon_{k+N_c}\left(\Xi\right)}{\ell_c \left(\Hat{x}_{k\vert k}, \hat{u}_{k\vert k} \right)} =&
     \frac{\Upsilon_{k+N_c} \left(\Xi\right)} {\ell_c(x_{k+N_c-1},u_{k+N_c-1})} \, \frac{\ell_c(x_{k+N_c-1},u_{k+N_c-1})}{\ell_c(x_{k+N_c-2},u_{k+N_c-2})}\,\ldots\,\frac{\ell_c(x_{k+1},u_{k+1})}{\ell_c(x_{k}),u_{k}},
\end{align}

The term $\Upsilon_{k+N_c}\left(\Xi\right)$ can be upper as follows

\begin{align}
    \Upsilon_{k+N_c}\left(\Xi\right) =& \Psi_{C,k,N_c} - \sum_{j=0}^{N_c-1} \ell_c(x_j,u_j), \\
    \leq& L_{N_c}\sigma(x)-\sigma(x), \\
    =& \sigma(x)(L_{N_c}-1),
\end{align}

Then, then term $\ell_c(x_{k+N_c-1},u_{k+N_c-1})$ is upper bounded as

\begin{align}
    \ell_c(x_{k+N_c-1},u_{k+N_c-1}) =&\, \Psi_{C,k,N_c} -\left(\sum_{j=0}^{N_c-2}\ell_c(x_j,u_j)+\Upsilon_{k+N_c}\left(\Xi\right)\right),\\
    \leq&\, L_{N_c}\sigma(x)-\sigma(x),\\
    =&\, \sigma(x)(L_{N_c}-1),
\end{align}
The term $\ell_c(x_{k+N_c-2},u_{k+N_c-2})$ is bounded by

\begin{align}
    \ell_c(x_{k+N_c-2},u_{k+N_c-2}) =&\, \Psi_{\textnormal{\fontsize{6pt}{8pt}\selectfont\textit{C,k,$N_c-1$}}}-\left(\sum_{j=0}^{N_c-3}\ell_c(x_j,u_j)-\Upsilon_{k+N_c-1}\left(x_{k+N_c-1}\right)\right),\\
    \leq&\, L_{N_c-1}\sigma(x)-\sigma(x),\\
    =&\, \sigma(x)(L_{N_c-1}-1),
\end{align}
The remaining terms are bounded similarly. Then, by noting that the terms in the denominators in \eqref{eq: prod of cocientes} delayed one sampling time respect the terms in the numerators, \eqref{eq: prod of cocientes} can be bounded as

\begin{align}
    \frac{\Upsilon_{k+N_c}\left(\Xi\right)}{\ell_c \left(\Hat{x}_{k\vert k}, \hat{u}_{k\vert k} \right)} \leq& \prod_{j=1}^{N_c}\frac{L_i-1}{L_{i-1}}\leq \prod_{j=1}^{N_c}\frac{L-1}{L}=\left(\frac{L-1}{L}\right)^{N_c}=\frac{(L-1)}{L}\left(\frac{L-1}{L}\right)^{N_c-1},\\
    \leq&\, (L-1)\left(\frac{L-1}{L}\right)^{N_c-1},
\end{align}
Then, Inequality \eqref{eq:condition for stability 2} can be stated as

\begin{align}
    \frac{\Upsilon_{k+N_c}\left(\Xi\right)}{\ell_c \left(\Hat{x}_{k\vert k}, \hat{u}_{k\vert k} \right)} \leq&\, (L-1)\left(\frac{L-1}{L}\right)^{N_c-1} \leq \frac{1-\Delta^w_c}{\delta},
\end{align}
Then, by algebraic manipulations, one can verify that if the length of the forward window is selected as

\begin{align}   
   N_c =& \ceil*{\displaystyle \frac{ \ln{\left(\frac{\delta\left(L-1\right)}{1-\bar{\Delta}^w_c}\right)}}{\ln{\left(\frac{L}{L-1}\right)}} + 1}.
\end{align}
Then, the stability condition \eqref{eq:condition for stability 2} is verified, and $\delta\omega < 1$.}
{\hfill$\square$}}

\newpage

\section{Derivation of $\Dot{p}_t$} \label{Deriv_p}

\begin{align}
    \Dot{p}_t =& \frac{\Delta x}{\vert \Delta x \vert}\left(\Dot{x}^{(1)}_t-\Dot{x}^{(2)}_t\right),\nonumber\\
              =& \frac{\Delta x}{\vert \Delta x \vert}\left(ax^{(1)^3}_t+w^{(1)}_t+u^{(1)}_t-ax^{(2)^3}_t-w^{(2)}_t-u^{(2)}_t\right),\nonumber \\
              =& \frac{\Delta x}{\vert \Delta x \vert}\left(a\left(x^{(1)^3}_t-x^{(2)^3}_t\right)-K\Delta x+\Delta w_t\right),\nonumber\\
              =& \frac{\Delta x}{\vert \Delta x \vert}\left(a\Delta x\left(x^{(1)^2}_t+x^{(1)}_t x^{(2)}_t+x^{(2)^2}_t\right)-K\Delta x+\Delta w_t\right),\nonumber\\
           \leq& -K\vert \Delta x \vert+\vert \Delta x \vert a \left(x^{(1)^2}_t+x^{(1)}_t x^{(2)}_t+x^{(2)^2}_t\right)+\vert \Delta w_t \vert,\nonumber\\
           \leq& -K\vert \Delta x \vert+\vert \Delta x \vert a\frac{\left(x^{(1)^3}-x^{(2)^3}\right)}{\Delta x}+\vert \Delta w_t \vert,\nonumber\\
    \leq& -Kp_t + a\left(\left(y^{(1)}_t-v^{(1)}_t\right)^3-\left(y^{(2)}_t-v^{(2)}_t\right)^3\right)+\vert \Delta w_t \vert,\nonumber\\
    \leq& -Kp_t + a\vert h^3\left(x^{(1)}_t\right)-h^3\left(x^{(2)}_t\right) \vert+\vert \Delta w_t \vert,\nonumber\\
    \leq& -Kp_t +ag\vert \Delta h_t \vert  +\vert\Delta w_t \vert.\nonumber
\end{align}

{\hfill$\square$}

\bibliography{wileyNJD-AMA}%

\clearpage

\end{document}